\documentclass[prc,preprint,
  showpacs,showkeys,lengthcheck,
  nofootinbib,tightenlines,onecolumn,notitlepage,
  preprintnumbers,superscriptaddress]{revtex4-1}

\usepackage{graphicx}
\usepackage[usenames,dvipsnames]{xcolor}
\graphicspath{{figures/}}

\usepackage{xr-hyper}
\usepackage{hyperref}
\hypersetup{colorlinks=true,citecolor=blue,linkcolor=blue,urlcolor=blue}
\usepackage{diagbox}
\usepackage{booktabs}

\usepackage{amsmath,amssymb,amsfonts}
\usepackage{mathrsfs}
\usepackage{slashed}
\usepackage{bm}
\usepackage{tensor}

\usepackage[final]{changes} 
\definechangesauthor[color=orange!76!black]{Adam}
\definechangesauthor[color=green!50!black]{Wim}
\definechangesauthor[color=blue]{rev}  

\begin{document}

\title{Spatial densities of momentum and forces in spin-one hadrons}

\author{Adam Freese}
\email{afreese@uw.edu}
\affiliation{Department of Physics, University of Washington, Seattle, WA 98195, USA}

\author{Wim Cosyn}
\email{wcosyn@fiu.edu}
\affiliation{Department of Physics, Florida International University, Miami, FL 33199, USA}
\affiliation{Department of Physics and Astronomy, Ghent University, B9000 Gent, Belgium}

\begin{abstract}
  Densities associated with the energy-momentum tensor are calculated for
  spin-one targets.
  These calculations are done in a light front formalism,
  which accounts for relativistic effects due to boosts
  and allows for arbitrary spatial localization of the target.
  These densities include the distribution of momentum, angular momentum,
  and pressures over a two-dimensional plane transverse to the light front.
  Results are obtained for both longitudinally and transversely polarized
  targets, and the formalism is tailored to allow the possibility of massless
  targets.
  The momentum density and pressure distributions are calculated for
  a deuteron target in a light cone convolution model,
  with which the properties of this model
  (such as helicity dependence of the densities)
  is illustrated.
\end{abstract}

\preprint{NT@UW-2206}

\maketitle



\section{Introduction}

The energy momentum tensor (EMT) has become a major topic of interest
in hadron physics.
It touches on several major outstanding problems in the field,
including the proton mass
puzzle~\cite{Ji:1994av,Ji:1995sv,Lorce:2017xzd,Hatta:2018sqd,Metz:2020vxd,Ji:2021mtz,Lorce:2021xku}
and the proton spin
puzzle~\cite{Ashman:1987hv,Ji:1996ek,Leader:2013jra}.
It is also believed by some to contain information about
the mechanical properties of hadrons,
including the spatial distributions of pressures and shear
stresses~\cite{Polyakov:2002yz,Perevalova:2016dln,Polyakov:2018zvc},
as well as information about the mechanical stability of hadrons.

Most research into the EMT of hadrons has focused on
the gravitational form factors (GFFs) of spin-zero and spin-half targets.
This is understandable, since the proton is spin-half,
and spin-zero is an especially simple case for exploratory studies.
However, spin-one targets play an important role in our understanding of
the strong nuclear force, and are thus deserving of more attention
in research on GFFs.
The deuteron is spin-one after all, and as the simplest nucleus,
it is an ideal testing ground for studies of
how the inter-nucleon force arises from quantum
chromodynamics~\cite{Boeglin:2015cha}.
Spin-one targets more generally contain extra information not present
in lower-spin targets,
such as a gluon transversity distribution whose evolution decouples from
quarks~\cite{Jaffe:1989xy}.

Several recent theoretical studies~\cite{Taneja:2011sy,Cosyn:2019aio,Polyakov:2019lbq,Kim:2022wkc}
and model calculations~\cite{Abidin:2008ku,Freese:2019bhb,Sun:2020wfo,Epelbaum:2021ahi}
have been done for the EMT and GFFs of spin-one targets.
However, there is yet no investigation into the light front densities
associated with the GFFs of spin-one targets.
Breit frame studies exist~\cite{Cosyn:2019aio,Polyakov:2019lbq},
but there is considerable controversy regarding the physical meaningfulness
of Breit frame densities
(see Refs.~\cite{Fleming:1974af,Burkardt:2002hr,Miller:2018ybm,Lorce:2018egm,Jaffe:2020ebz,Lorce:2020onh,Freese:2021czn,Epelbaum:2022fjc} for a variety of perspectives),
whereas light front densities have a clear physical meaning and interpretation
as true densities~\cite{Burkardt:2002hr,Miller:2007uy,Miller:2009sg,Miller:2018ybm,Freese:2021czn}.
It is thus prudent to investigate the light front densities associated
with the GFFs of spin-one targets.

This work is an investigation into the general properties and expressions
for EMT densities in spin-one targets.
A companion paper \cite{Freese:2022ibw}
investigates the densities for a photon target specifically.

This paper is organized into the following sections.
Sec.~\ref{sec:elements} considers the decomposition of EMT matrix elements
into GFFs, examining how this decomposition depends on target polarization.
Sec.~\ref{sec:densities} then obtains all the relevant densities,
including static moments and radii, as well as their polarization dependence.
Sec.~\ref{sec:example} illustrates some of these densities with a simple
light cone convolution model of the deuteron,
and Sec.~\ref{sec:conclusion} concludes the work.


\section{Matrix elements for definite-spin states}
\label{sec:elements}

For a massive spin-one system,
the matrix element of the conserved, symmetric EMT between
spin-one plane wave states is given
by~\cite{Holstein:2006ge,Abidin:2008ku,Taneja:2011sy,Cosyn:2019aio,Freese:2019bhb,Polyakov:2019lbq}:
\begin{align}
  \label{eqn:emt:massive}
  \langle p'\lambda' | T^{\mu\nu}(0) | p \lambda \rangle
  &=
  2 P^\mu P^\nu
  \left[
    -(\varepsilon\cdot\varepsilon'^*)
    \mathcal{G}_1(t)
    + \frac{(\varepsilon\cdot\Delta)(\varepsilon'^*\cdot\Delta)}{2M^2}
    \mathcal{G}_2(t)
    \right]
  \notag \\ &
  +
  \frac{\Delta^\mu\Delta^\nu-\Delta^2g^{\mu\nu}}{2}
  \left[
    -(\varepsilon\cdot\varepsilon'^*)
    \mathcal{G}_3(t)
    + \frac{(\varepsilon\cdot\Delta)(\varepsilon'^*\cdot\Delta)}{2M^2}
    \mathcal{G}_4(t)
    \right]
  +
  \frac{1}{2}
  P^{\{\mu}
  \Big[
    \varepsilon'^{*\nu\}} (\varepsilon\cdot\Delta)
    -
    \varepsilon^{\nu\}} (\varepsilon'^*\cdot\Delta)
    \Big]
  \mathcal{G}_5(t)
  \notag \\ &
  +
  \frac{1}{4}
  \Big[
    \Delta^{\{\mu}
    \Big(
      \varepsilon'^{*\nu\}} (\varepsilon\cdot\Delta)
      +
      \varepsilon^{\nu\}} (\varepsilon'^*\cdot\Delta)
      \Big)
    -
    \varepsilon^{\{\mu}\varepsilon'^{*\nu\}} \Delta^2
    -
    2 g^{\mu\nu}
    (\varepsilon\cdot\Delta)(\varepsilon'^*\cdot\Delta)
    \Big]
  \mathcal{G}_6(t)
  \,,
\end{align}
where $P = \frac{1}{2}\big(p+p'\big)$,
$\Delta = p'-p$,
$t = \Delta^2$,
where $\varepsilon$ is a polarization four-vector that
depends on momentum $p$ and spin quantum number $\lambda$,
and $\varepsilon'$ similarly on $p'$ and $\lambda'$,
and
where $\{\}$ denotes symmetrization without a factor $\frac{1}{2}$
(i.e., $a^{\{\mu}b^{\nu\}} = a^\mu b^\nu + a^\nu b^\mu$).
Note that several conventions exist in the literature for naming
the gravitational form factors.
We have here used the notation first found in Ref.~\cite{Taneja:2011sy}
and later adopted (and expanded) in Refs.~\cite{Cosyn:2019aio,Freese:2019bhb}.
Ref.~\cite{Polyakov:2019lbq} gives a comparison of the existing conventions.
Several non-conserved form factors, namely $\mathcal{G}_{7-9}(t)$,
also exist when examining the EMT contributions of a single parton flavor,
but in this work we examine only the total EMT, which is conserved.
The effects of non-conserved GFFs on partonic densities
are deferred to a future study.
Ref.~\cite{Cosyn:2019aio} additionally gives two more
form factors, $\mathcal{G}_{10,11}(t)$ for the asymmetric EMT,
but a consistent application of Noether's second theorem to obtain the EMT
has been shown to reproduce the symmetric Belinfante EMT
for QCD~\cite{Freese:2021jqs},
so we limit our attention to the symmetric EMT here.

Clearly, Eq.~(\ref{eqn:emt:massive}) is not applicable to massless systems,
due to the presence of factors $1/M^2$.
The presence of these factors is somewhat artificial;
standard form factor decompositions like Eq.~(\ref{eqn:emt:massive})
are designed so that
(1) the form factors are unitless and
(2) poles do not occur in the form factors nor in accompanying
Lorentz structures that are not present in the EMT matrix element.
Condition (2) precludes using factors of $1/t$ instead of $1/M^2$
to accompany, e.g., $\mathcal{G}_2(t)$.
However, if condition (1) is relaxed, one can write a variant of
Eq.~(\ref{eqn:emt:massive}) with no factors of $1/M^2$ present,
but several unitful
Lorentz scalar functions.

This work will examine light front densities of spin-one systems,
including massless systems such as the photon.
It is thus desirable to have a
breakdown into Lorentz scalar functions
that is
applicable to both massless and massive systems.
When considering light front densities in particular,
where $\Delta^+=0$ by virtue of integrating out $x^-$~\cite{Freese:2021czn},
the EMT matrix element can be decomposed as follows:
\begin{align}
  \label{eqn:emt:spin1}
  \langle p'\lambda' | T^{\mu\nu}(0) | p \lambda \rangle
  \bigg|_{\Delta^+=0}
  &=
  2 P^\mu P^\nu
  \mathcal{A}_{\lambda'\lambda}(\boldsymbol{\Delta}_\perp)
  -
  i
  \frac{ P^{\{\mu}\epsilon^{\nu\}P\Delta n} }{ (P\cdot n) }
  \mathcal{J}_{\lambda'\lambda}(\boldsymbol{\Delta}_\perp)
  +
  \frac{ \Delta^\mu \Delta^\nu - \Delta^2 g^{\mu\nu} }{2}
  \mathcal{D}_{\lambda'\lambda}(\boldsymbol{\Delta}_\perp)
  \notag \\ &
  +
  \frac{P^{\{\mu}n^{\nu\}}}{(P\cdot n)}
  \mathcal{E}_{\lambda'\lambda}(\boldsymbol{\Delta}_\perp)
  +
  \frac{n^\mu n^\nu}{(P\cdot n)^2}
  \mathcal{H}_{\lambda'\lambda}(\boldsymbol{\Delta}_\perp)
  +
  i
  \frac{ n^{\{\mu}\epsilon^{\nu\}P\Delta n} }{ (P\cdot n)^2 }
  \mathcal{K}_{\lambda'\lambda}(\boldsymbol{\Delta}_\perp)
  \,,
\end{align}
where $n$ is the lightlike four-vector that defines the light front coordinates,
i.e., such that $V\cdot n = V^+$ and
the decomposition was constructed to be invariant under scaling $n$ by a factor.
It should be remarked that the Lorentz scalar functions
$\mathcal{A}_{\lambda'\lambda}(\boldsymbol{\Delta}_\perp)$ etc.\ are
not proper form factors,
owing to their dependence on the initial and final target helicities,
but can more accurately be called helicity amplitudes.
\added{
It should also be noted that this decomposition is not defined when $P^+ = 0$,
which can occur in the massless case for plane waves in the $-z$ direction.
}
This decomposition has several unitful
helicity amplitudes,
namely
$\mathcal{E}$, $\mathcal{H}$ and $\mathcal{K}$,
with units GeV$^2$, GeV$^4$ and GeV$^2$, respectively.

It should be stressed that we do not propose the
helicity amplitudes
in
Eq.~(\ref{eqn:emt:spin1}) as a replacement for any of the existing conventions;
their utility lies specifically in
the ability to take light front Fourier transformations
of these 
helicity amplitudes
to obtain physically interpretable densities.
In the respect that Fourier transforms of these quantities
produce light front densities
(similarly to form factors for spin-zero and spin-half targets),
we will occasionally refer to the helicity amplitudes as
``effective form factors,''
but we stress that these quantities are not really proper form factors.

In the massive case, the six
helicity amplitudes
in Eq.~(\ref{eqn:emt:spin1})
are linear combinations of the
form factors
found in Eq.~(\ref{eqn:emt:massive}),
with the particular combination depending on the initial and final helicity.
Of special interest are those that contribute to the
Galilean densities~\cite{Lorce:2018egm,Freese:2021czn},
which are the densities corresponding to only the $+$
and transverse spatial components of the EMT.
These densities have the special property of being covariant
under the Galilean subgroup of the Poincar\'e group.
Since $n^+=0$ and $\mathbf{n}_\perp^i = 0$
(we use bold vectors with a $_\perp$ subscript to signify
transverse spatial components),
only $\mathcal{A}$, $\mathcal{J}$, and $\mathcal{D}$ contribute
to these densities.

The relationships between the form factors in
Eq.~(\ref{eqn:emt:massive}) and
helicity amplitudes in Eq.~(\ref{eqn:emt:spin1})
can be found by evaluating
Eq.~(\ref{eqn:emt:massive}) explicitly using the spin-one polarization vectors
found in Ref.~\cite{Berger:2001zb} at $\xi \equiv -\frac{\Delta^+}{2P^+}=0$
(which are also given in Appendix~\ref{app:polar}).
For example, let us consider cases with no helicity flip
($\lambda'=\lambda$),
which are relevant to the Galilean densities of light front
helicity states.
For helicity $\pm1$ states we have:
\begin{subequations}
  \label{eqn:gff:hel1}
  \begin{align}
    \mathcal{A}_{\pm\pm}(\boldsymbol{\Delta}_\perp)
    &=
    \mathcal{G}_1(t)
    - \frac{t}{4M^2} \mathcal{G}_2(t)
    \\
    \mathcal{J}_{\pm\pm}(\boldsymbol{\Delta}_\perp)
    &=
    \label{eqn:Jt}
    \pm\frac{1}{2}\mathcal{G}_5(t)
    \equiv
    \pm \mathscr{J}(t)
    \\
    \mathcal{D}_{\pm\pm}(\boldsymbol{\Delta}_\perp)
    &=
    \mathcal{G}_3(t)
    - \mathcal{G}_6(t)
    - \frac{t}{4M^2} \mathcal{G}_4(t)
    \,,
  \end{align}
\end{subequations}
whereas for helicity-0 states we have:
\begin{subequations}
  \label{eqn:gff:hel0}
  \begin{align}
    \mathcal{A}_{00}(\boldsymbol{\Delta}_\perp)
    &=
    \left( 1 + \frac{t}{2M^2} \right)
    \mathcal{G}_1(t)
    - \frac{t}{4M^2} \Big(
    2 \mathcal{G}_5(t) + \mathcal{G}_6(t)
    \Big)
    - \frac{t^2}{8M^4} \mathcal{G}_2(t)
    \\
    \mathcal{J}_{00}(\boldsymbol{\Delta}_\perp)
    &=
    0 \label{eqn:gff:hel0:J}
    \\
    \mathcal{D}_{00}(\boldsymbol{\Delta}_\perp)
    &=
    \mathcal{G}_3(t)
    - \frac{t}{4M^2} \Big(
    -2 \mathcal{G}_3(t) + \mathcal{G}_6(t)
    \Big)
    - \frac{t^2}{8M^4} \mathcal{G}_4(t)
    \,.
  \end{align}
\end{subequations}
The results for the other
helicity amplitudes
can be found in Appendix~\ref{app:elements}.

\subsection{Transversely polarized states}

We next consider transversely polarized states for \emph{massive}
spin-one hadrons.
Since the only sensible quantization axis for the spin of massless particles
is along the direction of travel,
transversely polarized states can only sensibly be considered in
the massive case.
The transverse polarization vectors are given by the following linear
combinations of light front helicity states~\cite{Carlson:2008zc}:
\begin{subequations}
  \label{eqn:polar:trans}
  \begin{align}
    \varepsilon_{T,\pm1}^\mu
    &=
    \frac{ \varepsilon_{+1}
    \pm\sqrt{2}e^{i\phi_s}\varepsilon_0
    + e^{2i\phi_s}\varepsilon_{-1} }{2}
    \\
    \varepsilon_{T,0}^\mu
    &=
    \frac{ \varepsilon_{+1} - e^{2i\phi_s}\varepsilon_{-1} }{\sqrt{2}}
    \,,
  \end{align}
\end{subequations}
and similarly for the final (primed) state.
Accordingly, the relevant EMT matrix elements will involve spin-flip
contributions.
A catalogue of all the individual contributions can be found in
Appendix~\ref{app:elements}.
Without loss of generality, we can define $\hat{x}=\mathbf{s}_\perp$,
and for more compact formulas, we suppress explicit $\mathbf{s}_\perp$
dependence in the expressions to follow.

The simplest manner to give results is in terms of the
effective form factors:
\begin{align}
  \label{eqn:emt:spin1:trans}
  \langle p',m_s| T^{\mu\nu}(0) | p,m_s\rangle
  &=
  2 P^\mu P^\nu
  \mathcal{A}_T^{(m_s)}(\boldsymbol{\Delta}_\perp)
  -
  i
  \frac{ P^{\{\mu}\epsilon^{\nu\}P\Delta n} }{ (P\cdot n) }
  \mathcal{J}_T^{(m_s)}(\boldsymbol{\Delta}_\perp)
  +
  \frac{ \Delta^\mu \Delta^\nu - \Delta^2 g^{\mu\nu} }{2}
  \mathcal{D}_T^{(m_s)}(\boldsymbol{\Delta}_\perp)
  \notag \\ &
  +
  \frac{P^{\{\mu}n^{\nu\}}}{(P\cdot n)}
  \mathcal{E}_T^{(m_s)}(\boldsymbol{\Delta}_\perp)
  +
  \frac{n^\mu n^\nu}{(P\cdot n)^2}
  \mathcal{H}_T^{(m_s)}(\boldsymbol{\Delta}_\perp)
  +
  i
  \frac{ n^{\{\mu}\epsilon^{\nu\}P\Delta n} }{ (P\cdot n)^2 }
  \mathcal{K}_T^{(m_s)}(\boldsymbol{\Delta}_\perp)
  \,,
\end{align}
where $m_s \in \{-1,0,+1\}$ is the magnetic spin number, i.e.,
the eigenvalue of $\mathbf{s}_\perp$ projected along the quantization axis.
Each of the effective form factors works out to have the form:
\begin{subequations}
  \label{eqn:azimuthal}
  \begin{align}
    \mathcal{F}^{(\pm1)}_T(\boldsymbol{\Delta}_\perp)
    &=
    \frac{1}{4}\Big(
    \mathcal{F}_{++}(t)
    +
    \mathcal{F}_{--}(t)
    +
    2
    \mathcal{F}_{00}(t)
    \Big)
    +
    \frac{t\cos 2\phi_\Delta }{8M^2}
    \,
    \mathcal{F}^{\cos 2\phi }_T(t)
    \pm
    \frac{i\sqrt{-t}\sin \phi_\Delta }{2M}
    \mathcal{F}^{\sin \phi }_T(t)
    \\
    \mathcal{F}^{(0)}_T(\boldsymbol{\Delta}_\perp)
    &=
    \frac{1}{2}\Big(
    \mathcal{F}_{++}(t)
    +
    \mathcal{F}_{--}(t)
    \Big)
    -
    \frac{t\cos 2\phi_\Delta }{4M^2}
    \,
    \mathcal{F}^{\cos 2\phi }_T(t)
    \,,
  \end{align}
\end{subequations}
where $\mathcal{F}$ stands in for any of the effective form factors
in Eq.~(\ref{eqn:emt:spin1}) or Eq.~(\ref{eqn:emt:spin1:trans}),
and where $\phi_\Delta$ is the angle between $\boldsymbol{\Delta}_\perp$
and $\mathbf{s}_\perp$.
The $\mathcal{F}^{\sin \phi }_T(t)$ and $\mathcal{F}^{\cos 2\phi }_T(t)$
that are relevant to the Galilean light front densities are,
in terms of the traditional GFFs:
\begin{subequations}
  \label{eqn:AJD:mod}
  \begin{align}
    \mathcal{A}^{\sin\phi}_T(t)
    &=
    \mathcal{G}_5(t)
    -
    2\mathcal{G}_1(t)
    +
    \frac{t}{2M^2} \mathcal{G}_2(t)
    \\
    \mathcal{A}^{\cos 2\phi}_T(t)
    &=
    \mathcal{G}_2(t)
    \\
    \mathcal{J}^{\sin\phi}_T(t)
    &=
    0
    \\
    \mathcal{J}^{\cos 2\phi}_T(t)
    &=
    0
    \\
    \mathcal{D}^{\sin\phi}_T(t)
    &=
    \mathcal{G}_6(t)
    -
    2 \mathcal{G}_3(t)
    +
    \frac{t}{2M^2} \mathcal{G}_4(t)
    \\
    \mathcal{D}^{\cos 2\phi}_T(t)
    &=
    \mathcal{G}_4(t)
    \,.
  \end{align}
\end{subequations}
It should be clear from these results,
combined with Eqs.~(\ref{eqn:Jt}) and (\ref{eqn:gff:hel0:J}),
that $\mathcal{J}_T=0$.
This means that the
$J_z$
density for transversely polarized states is zero, see Eq.~(\ref{eqn:J:hel}) further down.
To be sure, the structure of Eq.~(\ref{eqn:emt:spin1}) means that
$\mathcal{J}_T$ is a $J_z$ density of transversely polarized states
and not a density of transverse angular momentum.
This
finding is not surprising,
since the expectation value of $J_z$ in a
transversely polarized state is zero.
The transverse angular momentum operator is given by
$x^- T^{+i}(x) - x_\perp^i T^{+-}(x)$,
meaning that this would entail a non-Galilean density
and is thus beyond the scope of this work.
Indeed, this quantity would involve several of the non-Galilean
helicity amplitudes (those other than $\mathcal{A}$, $\mathcal{J}$
and $\mathcal{D}$).

It may be instructive to consider Eq.~(\ref{eqn:azimuthal})
in terms of unpolarized, vector polarized, and tensor polarized combinations of the target.
These polarization combinations are defined for transversely polarized targets as follows:
\begin{subequations}
  \begin{align}
    \mathcal{F}_T^{(U)}(\boldsymbol{\Delta}_\perp)
    &=
    \frac{1}{3}
    \Big(
    \mathcal{F}^{(+1)}_T(\boldsymbol{\Delta}_\perp)
    +
    \mathcal{F}^{(-1)}_T(\boldsymbol{\Delta}_\perp)
    +
    \mathcal{F}^{(0)}_T(\boldsymbol{\Delta}_\perp)
    \Big)
    \\
    \mathcal{F}_T^{(V)}(\boldsymbol{\Delta}_\perp)
    &=
    \mathcal{F}^{(+1)}_T(\boldsymbol{\Delta}_\perp)
    -
    \mathcal{F}^{(-1)}_T(\boldsymbol{\Delta}_\perp)
    \\
    \mathcal{F}_T^{(T)}(\boldsymbol{\Delta}_\perp)
    &=
    \frac{1}{2}
    \Big(
    2
    \mathcal{F}^{(0)}_T(\boldsymbol{\Delta}_\perp)
    -
    \mathcal{F}^{(+1)}_T(\boldsymbol{\Delta}_\perp)
    -
    \mathcal{F}^{(-1)}_T(\boldsymbol{\Delta}_\perp)
    \Big)
    \,.
  \end{align}
\end{subequations}
If we also define these states for longitudinally polarized targets:
\begin{subequations}
  \begin{align}
    \mathcal{F}_L^{(U)}(t)
    &=
    \frac{1}{3}
    \Big(
    \mathcal{F}_{++}(t)
    +
    \mathcal{F}_{--}(t)
    +
    \mathcal{F}_{00}(t)
    \Big)
    \\
    \mathcal{F}_L^{(V)}(t)
    &=
    \mathcal{F}_{++}(t)
    -
    \mathcal{F}_{--}(t)
    \\
    \mathcal{F}_L^{(T)}(t)
    &=
    \frac{1}{2}
    \Big(
    2
    \mathcal{F}_{00}(t)
    -
    \mathcal{F}_{++}(t)
    -
    \mathcal{F}_{--}(t)
    \Big)
    \,,
  \end{align}
\end{subequations}
then we find for the transversely polarized states that:
\begin{subequations}
  \begin{align}
    \mathcal{F}_T^{(U)}(\boldsymbol{\Delta}_\perp)
    &=
    \mathcal{F}_L^{(U)}(t)
    \\
    \mathcal{F}_T^{(V)}(\boldsymbol{\Delta}_\perp)
    &=
    \frac{i\sqrt{-t}\sin \phi_\Delta }{M}
    \mathcal{F}^{\sin \phi }_T(t)
    \\
    \mathcal{F}_T^{(T)}(\boldsymbol{\Delta}_\perp)
    &=
    -
    \mathcal{F}_L^{(T)}(t)
    -
    \frac{t\cos 2\phi_\Delta }{4M^2}
    \,
    \mathcal{F}^{\cos 2\phi }_T(t)
    \,.
  \end{align}
\end{subequations}
The modulations can thus be interpreted in terms of vector and tensor polarization states,
but these states actually differ depending on the spin quantization axis.
We will consider general polarization below.
Throughout the remainder of this work, however,
we focus on deuterons in specific polarization states rather than mixtures.
The reason for this is that unpolarized, vector polarized, and tensor polarized states are
mixtures that are not present in the Hilbert space of the target,
and we choose to focus in this work on the densities and properties of spin-one systems
in pure states.

\subsection{General polarization}

An alternate way of considering the dependence on the initial and final state helicities of the spin-1 particle in Eq.~(\ref{eqn:emt:spin1}) is by tracing it with a spin-1 density matrix $\rho(\lambda,\lambda')$ characterizing the ensemble:
\begin{equation}
    \langle\langle  T^{\mu\nu}(0) \rangle\rangle \equiv  \sum_{\lambda,\lambda'} \rho(\lambda,\lambda') \langle p'\lambda' | T^{\mu\nu}(0) | p \lambda \rangle .
\end{equation}
We refer the reader to Appendix~\ref{app:rho} for a summary of the spin-1 density matrix formalism and a definition of the density matrix parameters ($S_L, S_T, \phi_S, T_{LL}, T_{LT}, T_{TT}, \phi_{T_L}, \phi_{T_T}$) appearing in the formulas that follow.  By considering the relevant contractions of the off-diagonal covariant density matrix of Eq.~(\ref{eqn:rho_off}), evaluated at $\Delta^+=0$, with tensors built from $P,\Delta,n$ and the 4D Levi-Civita tensor, we obtain the following expressions for the density matrix averaged effective form factors appearing in Eq.~(\ref{eqn:emt:spin1}):

\begin{subequations}
\label{eqn:gff_genpol}
\begin{align}
    \sum_{\lambda,\lambda'} \rho(\lambda,\lambda')\, \mathcal{A}_{\lambda' \lambda} =&
    \left(\frac{2}{3}+T_{LL}\right)\left(\mathcal{G}_1(t) - \frac{t}{4M^2}\mathcal{G}_2(t) \right)+ \left(\frac{1}{3}-T_{LL}\right)\left[\left(1+\frac{t}{2M^2}\right) \mathcal{G}_1(t)-\frac{t^2}{8M^4}\mathcal{G}_2(t)\right.\nonumber\\ 
    & \qquad\left. -\frac{t}{2M^2}\mathcal{G}_5(t)-\frac{t}{4M^2}\mathcal{G}_6(t) \right] 
     \,+\, iS_T\sin(\phi_S-\phi_t)\, \frac{\sqrt{-t}}{M}\left(\mathcal{G}_1(t)-\frac{t}{4M^2}\mathcal{G}_2(t)-\frac{\mathcal{G}_5(t)}{2} \right) \nonumber\\
     & \qquad + T_{TT}\cos(2\phi_{T_T}-2\phi_t)\, \frac{t}{4M^2}\mathcal{G}_2(t)\,,
    \\[1em]
    \sum_{\lambda,\lambda'} \rho(\lambda,\lambda')\, \mathcal{J}_{\lambda' \lambda} =& \frac{S_L}{2}\,\mathcal{G}_5(t) + i\,T_{LT}\sin(\phi_{T_L}-\phi_t)\frac{\sqrt{-t}}{2M}\left( \mathcal{G}_5(t)+\mathcal{G}_6(t)\right)\,,\\[1em]
    \sum_{\lambda,\lambda'} \rho(\lambda,\lambda')\, \mathcal{D}_{\lambda' \lambda} =&
    \left(\frac{2}{3}+T_{LL}\right)\left(\mathcal{G}_3(t)-\frac{t}{4M^2}\mathcal{G}_4(t)-\mathcal{G}_6(t) \right) +\left(\frac{1}{3}-T_{LL}\right)\left[\left(1+\frac{t}{2M^2}\right)\mathcal{G}_3(t) \right.\nonumber\\
    &\qquad \left. - \frac{t^2}{8M^4}\mathcal{G}_4(t)-\frac{t}{4M^2}\mathcal{G}_6(t) \right]
     \,+\, iS_T \sin(\phi_S - \phi_t) \, \frac{\sqrt{-t}}{M}\left(\mathcal{G}_3(t)-\frac{t}{4M^2}\mathcal{G}_4(t)-\frac{\mathcal{G}_6(t)}{2} \right)\nonumber\\
     &\qquad + T_{TT}\cos(2\phi_{T_T}-2\phi_t)\, \frac{t}{4M^2}\mathcal{G}_4(t)\,,
    \end{align}
\end{subequations}
where $\phi_t$ is the azimuthal angle of $\bm{\Delta}_\perp$.

We point out that $\mathcal{A}$ and $\mathcal{D}$ depend on the same density matrix parameters---namely,
$1$ (the unpolarized part), $T_{LL}, S_T\sin(\phi_S-\phi_t)$, and $T_{TT}\cos(2\phi_{T_T}-2\phi_t)$---while
$\mathcal{J}$ depends on $S_L$ and $T_{LT}\sin(\phi_{T_L}-\phi_t)$.
The first set of parameters are scalars and the second set are pseudoscalars,
reflecting the parity properties of the central charge $p^+$ and pressure on the one hand,
and particle spin on the other.
The modulations in the densities for the transversely polarized states are identified with the transverse vector ($S_T\sin\phi$) and mixed longitudinal-transverse tensor  ($T_{LT}\sin\phi$) polarized part of the density matrix for the $\sin\phi$ modulations, and the completely transverse tensor part ($T_{TT}\cos2\phi$) for the $\cos2\phi$ modulations.

For Eq.~(\ref{eqn:gff_genpol}), our previous expressions for the helicity [Eqs.~(\ref{eqn:gff:hel1}) and (\ref{eqn:gff:hel0})] and transversely polarized states [Eqs.~(\ref{eqn:azimuthal}) and (\ref{eqn:AJD:mod})] are recovered after identifying the corresponding density matrix parameters for these ensembles.  For pure longitudinal polarized states we need the following rest frame spin parameters
\begin{subequations}
\begin{align}
    &\lambda = \pm 1: &S_L = \pm1,\; T_{LL}=1/3,\nonumber\\
    && S_T=T_{LT} = T_{TT}= 0,\\[1em]
    &\lambda = 0:  &T_{LL}=-2/3, \nonumber\\
    &&S_L=S_T=T_{LT} = T_{TT}= 0.
\end{align}
\end{subequations}
For pure transversely polarized states one has
\begin{subequations}
\begin{align}
    &\lambda = \pm 1: &S_T=\pm 1,\; T_{LL}=-1/6,\; T_{TT}=1/2,\; \phi_{T_T}=\phi_S,\nonumber\\
    && S_L= T_{LT} = 0,\\[1em]
    &\lambda = 0:  &T_{LL}=1/3,\; T_{TT}=-1,\; \phi_{T_T}=\phi_S,\nonumber\\
    &&S_T=S_L= T_{LT} = 0.
\end{align}
\end{subequations}


\section{Properties of spin one densities}
\label{sec:densities}

For states with definite light front helicity,
the effective form factors as given in Eqs.~(\ref{eqn:gff:hel1}) and (\ref{eqn:gff:hel0})
can be used to obtain the azimuthally symmetric
light front $P^+$, angular momentum, and pressure densities
of a spin-one system localized in the transverse plane.
The formulas for the $P^+$ density and comoving stress tensor
are identical to those already found
in Refs.~\cite{Lorce:2018egm,Freese:2021czn}
for spin-zero or spin-half hadrons,
but with $\mathcal{A}$ and $\mathcal{D}$ substituted for $A$ and $D$.
We give these relations again here (along with the angular momentum density):
\begin{subequations}
  \begin{align}
    \label{eqn:p+:hel}
    \rho_{p^+}^{(\lambda)}(b_\perp)
    &=
    P^+
    \int \frac{\mathrm{d}^2\boldsymbol{\Delta}_\perp}{(2\pi)^2}
    \mathcal{A}_{\lambda\lambda}(t)
    e^{-i\boldsymbol{\Delta}_\perp\cdot\mathbf{b}_\perp}
    \\
    \label{eqn:J:hel}
    \rho_{J_z}^{(\lambda)}(b_\perp)
    &=
    \lambda
    \int \frac{\mathrm{d}^2\boldsymbol{\Delta}_\perp}{(2\pi)^2}
    \left\{
      \mathcal{J}(t)
      +
      t \frac{\mathrm{d} \mathcal{J}(t)}{\mathrm{d}t}
      \right\}
    e^{-i\boldsymbol{\Delta}_\perp\cdot\mathbf{b}_\perp}
    \\
    \label{eqn:Sij:hel}
    S_\lambda^{ij}(\mathbf{b}_\perp)
    &=
    \frac{1}{4P^+}
    \int \frac{\mathrm{d}^2\boldsymbol{\Delta}_\perp}{(2\pi)^2}
    \Big(
    \boldsymbol{\Delta}_\perp^i \boldsymbol{\Delta}_\perp^j
    -
    \boldsymbol{\Delta}_\perp^2 \delta_{ij}
    \Big)
    \mathcal{D}_{\lambda\lambda}(t)
    e^{-i\boldsymbol{\Delta}_\perp\cdot\mathbf{b}_\perp}
    \,.
  \end{align}
\end{subequations}
The only significant difference from
the spin-zero and spin-half cases is that the densities now depend
on $\lambda$,
meaning that
the distribution of momentum, angular momentum, and forces
will differ between spin-one hadrons of the same species
that are prepared in different helicity states.

For transversely polarized states of massive hadrons,
the effective form factors have azimuthal dependence.
The exact manner of this dependence varies between the
densities under consideration, so we will
proceed to consider the properties of each separately.


\subsection{Light front momentum density}

The $P^+$ density for helicity states is given already
by Eq.~(\ref{eqn:p+:hel}).
The transversely polarized $P^+$ density
contains azimuthal dependence which is essentially carried
over from the azimuthal dependence of the effective form factors,
since:
\begin{subequations}
  \begin{align}
    \int \frac{\mathrm{d}^2\boldsymbol{\Delta}_\perp}{(2\pi)^2}
    \frac{i\sqrt{-t}\sin \phi_\Delta }{2M}
    \mathcal{A}^{\sin \phi }_T(t)
    e^{-i\boldsymbol{\Delta}_\perp\cdot\mathbf{b}_\perp}
    &=
    -
    \frac{\sin \phi }{2M}
    \frac{\mathrm{d}}{\mathrm{d}b_\perp}
    \int \frac{\mathrm{d}^2\boldsymbol{\Delta}_\perp}{(2\pi)^2}
    \mathcal{A}^{\sin \phi }_T(t)
    e^{-i\boldsymbol{\Delta}_\perp\cdot\mathbf{b}_\perp}
    \\
    \int \frac{\mathrm{d}^2\boldsymbol{\Delta}_\perp}{(2\pi)^2}
    \frac{t\cos 2\phi_\Delta }{4M^2}
    \mathcal{A}^{\cos 2\phi }_T(t)
    e^{-i\boldsymbol{\Delta}_\perp\cdot\mathbf{b}_\perp}
    &=
    \frac{\cos 2\phi }{4M^2}
    \left\{
      \frac{\mathrm{d}^2}{\mathrm{d}b_\perp^2}
      -
      \frac{1}{b_\perp} \frac{\mathrm{d}}{\mathrm{d}b_\perp}
      \right\}
    \int \frac{\mathrm{d}^2\boldsymbol{\Delta}_\perp}{(2\pi)^2}
    \mathcal{A}^{\cos 2\phi }_T(t)
    e^{-i\boldsymbol{\Delta}_\perp\cdot\mathbf{b}_\perp}
    \,,
  \end{align}
\end{subequations}
where on the right hand side,
$\phi$ is the angle between $\mathbf{b}_\perp$ and $\mathbf{s}_\perp$.
In numerical applications, formulas involving derivatives may not be stable,
and it may be more helpful to use Hankel transforms instead:
\begin{subequations}
  \begin{align}
    \int \frac{\mathrm{d}^2\boldsymbol{\Delta}_\perp}{(2\pi)^2}
    \frac{i\sqrt{-t}\sin \phi_\Delta }{2M}
    \mathcal{A}^{\sin \phi }_T(t)
    e^{-i\boldsymbol{\Delta}_\perp\cdot\mathbf{b}_\perp}
    &=
    \frac{\sin\phi}{2M}
    \frac{1}{2\pi}
    \mathscr{H}_1\big[ k\mathcal{A}_T^{\sin\phi}(t=-k^2) \big](b_\perp)
    \\
    \int \frac{\mathrm{d}^2\boldsymbol{\Delta}_\perp}{(2\pi)^2}
    \frac{t\cos 2\phi_\Delta }{4M^2}
    \mathcal{A}^{\cos 2\phi }_T(t)
    e^{-i\boldsymbol{\Delta}_\perp\cdot\mathbf{b}_\perp}
    &=
    \frac{\cos 2\phi}{4M^2}
    \frac{1}{2\pi}
    \mathscr{H}_2\big[ k^2\mathcal{A}_T^{\cos 2\phi}(t=-k^2) \big](b_\perp)
  \end{align}
\end{subequations}
where the Hankel transform of order $\nu$ is defined by~\cite{Poularikas:2018trans}:
\begin{align}
  \mathscr{H}_\nu\big[ F(k) \big](b)
  &=
  \int_0^\infty \mathrm{d}k \, k J_\nu(bk) F(k)
  \,,
\end{align}
and where $J_\nu(x)$ is the Bessel function of order $\nu$.

For the sake of more compact formulas,
it is prudent to define:
\begin{subequations}
  \label{eqn:hankel:A}
  \begin{align}
    \rho_{T}^{\sin \phi }(b_\perp)
    &=
    \frac{P^+}{2\pi}
    \frac{1}{2M}
    \mathscr{H}_1\big[ k\mathcal{A}_T^{\sin\phi}(-k^2) \big](b_\perp)
    \\
    \rho_{T}^{\cos 2\phi }(b_\perp)
    &=
    \frac{P^+}{2\pi}
    \frac{1}{4M^2}
    \mathscr{H}_2\big[ k^2\mathcal{A}_T^{\cos 2\phi}(-k^2) \big](b_\perp)
    \,.
  \end{align}
\end{subequations}
The $P^+$ density of a transversely polarized spin-one hadron
is thus given by:
\begin{subequations}
  \label{eqn:p+:trans}
  \begin{align}
    \rho_{T}^{(\pm1)}(\mathbf{b}_\perp)
    &=
    \frac{
      \rho_{p^+}^{(+)}(b_\perp) + \rho_{p^+}^{(0)}(b_\perp)
    }{2}
    \pm
    \sin \phi \,
    \rho_{T}^{\sin \phi }(b_\perp)
    +
    \frac{1}{2} \cos 2\phi \,
    \rho_{T}^{\cos 2\phi }(b_\perp)
    \\
    \rho_{T}^{(0)}(\mathbf{b}_\perp)
    &=
    \rho_{p^+}^{(+)}(b_\perp)
    -
    \cos 2\phi \,
    \rho_{T}^{\cos 2\phi }(b_\perp)
  \end{align}
\end{subequations}
where we have used
$\rho_{p^+}^{(+)}(b_\perp) = \rho_{p^+}^{(-)}(b_\perp)$
to make the formulas slightly shorter.

The $P^+$ densities for all polarization states satisfy sum rules.
Integrating Eq.~(\ref{eqn:p+:hel}) over all space gives:
\begin{align}
  \int \mathrm{d}^2\mathbf{b}_\perp \,
  \rho_{p^+}^{(\lambda)}(b_\perp)
  =
  P^+
  \mathcal{A}_{\lambda\lambda}(0)
  \,.
\end{align}
For this to equal $P^+$, we have the sum rule:
\begin{align}
  \mathcal{A}_{\lambda\lambda}(0)
  =
  1
  \,,
\end{align}
for each helicity $\lambda$.
Since $\mathcal{A}(0) = \mathcal{G}_1(0)$ for massive hadrons,
this is compatible with the $\mathcal{G}_1(0)=1$ sum rule
of Ref.~\cite{Cosyn:2019aio}.
Since the integrals of $\sin\phi$ and $\cos 2\phi$ over $[0,2\pi)$ are zero,
the azimuthal dependence of
$\rho_T^{(m_s)}(\mathbf{b}_\perp,\mathbf{s}_\perp)$
integrates to zero, and we also have:
\begin{align}
  \int \mathrm{d}^2\mathbf{b}_\perp \,
  \rho_T^{(m_s)}(\mathbf{b}_\perp,\mathbf{s}_\perp)
  =
  P^+
  \,.
\end{align}

The $P^+$ density for both helicity and transversely polarized states
do not have $P^+$ dipole moments, i.e.,
their center-of-$P^+$ is at the origin, as expected.
A general explanation for why this occurs can be found in Sec.~7
of Ref.~\cite{Lorce:2018zpf}.
For the case of helicity states, it is easy to see that:
\begin{align}
  \int \mathrm{d}^2\mathbf{b}_\perp \,
  \mathbf{b}_\perp
  \rho_{p^+}^{(\lambda)}(b_\perp)
  =
  0
  \,.
\end{align}
For transversely polarized states,
if we use coordinates where:
$\mathbf{s}_\perp = \hat{x}$:
\begin{align}
  \int \mathrm{d}^2\mathbf{b}_\perp \,
  \mathbf{b}_\perp
  \rho_T^{(m_s)}(\mathbf{b}_\perp,\mathbf{s}_\perp)
  =
  \frac{ m_s \hat{y} P^+ }{2M}
  \Big( \mathcal{G}_5(0) - 2\mathcal{G}_1(0) \Big)
  =
  0
  \,.
\end{align}
We know $\mathcal{G}_1(0) = 1$ by momentum conservation.
It has been shown previously~\cite{Abidin:2008ku,Taneja:2011sy,Cosyn:2019aio}
that $\mathcal{G}_5(0)=2$ follows from angular momentum
conservation.
Thus, the center-of-$P^+$ is at the origin, as expected.

On the other hand,
the transversely polarized $P^+$ density does exhibit a quadrupole moment.
In two spatial dimensions, we define the traceless quadrupole tensor as:
\begin{align}
  \mathcal{Q}_{\mathrm{LF}}^{ij}(\mathbf{s}_\perp,m_s)
  =
  \int \mathrm{d}^2\mathbf{b}_\perp \,
  ( 2 b_\perp^i b_\perp^j - b_\perp^2 \delta_{ij} )
  \rho_T^{(m_s)}(\mathbf{b}_\perp,\mathbf{s}_\perp)
  \,.
\end{align}
The quadrupole moment itself can be identified with:
\begin{align}
  \label{eqn:QLF}
  \mathcal{Q}_{\mathrm{LF}}(\mathbf{s}_\perp,m_s)
  =
  s_\perp^i s_\perp^j
  \mathcal{Q}_{\mathrm{LF}}^{ij}(\mathbf{s}_\perp,m_s)
  \,,
\end{align}
so that, conversely:
\begin{align}
  \mathcal{Q}_{\mathrm{LF}}^{ij}(\mathbf{s}_\perp,m_s)
  =
  \big( 2 s_\perp^i s_\perp^j - \delta_{ij} \big)
  \mathcal{Q}_{\mathrm{LF}}(\mathbf{s}_\perp,m_s)
  \,.
\end{align}
We find through explicit evaluation that:
\begin{subequations}
  \label{eqn:quadrupole}
  \begin{align}
    \mathcal{Q}_{\mathrm{LF}}(\mathbf{s}_\perp,\pm1)
    &=
    \frac{P^+}{2M^2}
    \mathcal{G}_2(0)
    \equiv
    \frac{1}{2} \mathcal{Q}_{\mathrm{LF}}
    \\
    \mathcal{Q}_{\mathrm{LF}}(\mathbf{s}_\perp,0)
    &=
    - \frac{P^+}{M^2}
    \mathcal{G}_2(0)
    \equiv
    - \mathcal{Q}_{\mathrm{LF}}
    \,.
  \end{align}
\end{subequations}
The value of $\mathcal{G}_2(0)$ is not constrained by any conservation
laws or sum rules, and is zero for a free boson~\cite{Polyakov:2019lbq}.
A non-zero quadrupole moment must thus be generated by dynamics.

As is conventional in the nuclear physics literature~\cite{Kellogg:1939zz},
a positive quadrupole moment indicates a prolate hadron
(elongated in the direction of the spin quantization axis),
while a negative quadrupole moment indicates an oblate hadron
(flattened in the direction of the spin axis).
A positive $\mathcal{G}_2(0)$ would thus mean that the
$m_s=\pm1$ state is prolate, and that the $m_s=0$ state is oblate.
A negative value for $\mathcal{G}_2(0)$ would of course
indicate the opposite.

The contrast with the Breit frame mass quadrupole moment
(see Refs.~\cite{Cosyn:2019aio,Polyakov:2019lbq})
is remarkable.
Comparing to Ref.~\cite{Cosyn:2019aio} in particular\footnote{
  The sign convention in
  Ref.~\cite{Freese:2019bhb}
  is the opposite as in Ref.~\cite{Cosyn:2019aio},
  the latter of which we follow in this work.
}
and using sum-rule enforced values
(and dropping non-conserved form factors):
\begin{align}
  \mathcal{Q}_{\mathrm{Breit}}
  =
  \frac{1}{M} \left\{
    \mathcal{G}_2(0)
    - 1 - \frac{1}{2}\mathcal{G}_6(0)
    \right\}
  \,.
\end{align}
The Breit frame quadrupole moment depends on
$\mathcal{G}_6(0)$ in addition to $\mathcal{G}_2(0)$.
Remarkably, $\mathcal{G}_6(0) = -2$ in the free theory~\cite{Polyakov:2019lbq},
meaning that the Breit frame mass quadrupole moment
is also generated entirely by dynamics.
However, since $\mathcal{G}_6(0)$ also comes into play,
the quadrupole moment may turn out to have different magnitudes
and even signs in the Breit frame and on the light front.
In Ref.~\cite{Freese:2019bhb},
the rho meson was found to have
$\mathcal{G}^{\mathrm{NJL}}_2(0) \approx 0.158 > 0$
and
$\mathcal{G}^{\mathrm{NJL}}_6(0) \approx -1$,
which means that the rho meson (in this model)
has a positive $P^+$ quadrupole moment on the light front,
but a negative mass quadrupole moment in the Breit frame.
Since $\mathcal{G}_6(0)$ is involved in the Breit frame quadrupole moment,
the difference between this and the light front quadrupole moment
may be due to relativistic spin effects,
as was remarked for the electric quadrupole moment in Ref.~\cite{Lorce:2022jyi}.

Let us lastly look at the $P^+$ radius of spin-one hadrons,
which is defined through:
\begin{align}
  \label{eqn:p+:radius}
  \langle b_\perp^2 \rangle_{P^+}
  =
  \frac{1}{P^+}
  \int \mathrm{d}^2\mathbf{b}_\perp \,
  \mathbf{b}_\perp^2
  \rho_{p^+}(\mathbf{b}_\perp)
  =
  4
  \frac{\partial\mathcal{A}(\boldsymbol{\Delta_\perp})}{\partial t}
  \bigg|_{t=0}
  \,,
\end{align}
and for massive hadrons differs between polarization states,
since the effective $\mathcal{A}(\boldsymbol{\Delta_\perp})$
form factor differs.
For helicity states of massive hadrons:
\begin{subequations}
  \begin{align}
    \langle b_\perp^2 \rangle_{P^+}(\lambda=\pm1)
    &=
    4
    \frac{\mathrm{d}\mathcal{G}_1(t)}{\mathrm{d}t}
    \bigg|_{t=0}
    -
    \frac{\mathcal{G}_2(0)}{M^2}
    \\
    \langle b_\perp^2 \rangle_{P^+}(\lambda=0)
    &=
    4
    \frac{\mathrm{d}\mathcal{G}_1(t)}{\mathrm{d}t}
    \bigg|_{t=0}
    +
    \frac{1}{M^2}
    \Big(
    2 \mathcal{G}_1(0) - 2 \mathcal{G}_5(0) - \mathcal{G}_6(0)
    \Big)
    \,,
  \end{align}
  while for transversely polarized states:
  \begin{align}
    \langle b_\perp^2 \rangle_{P^+}(m_s=\pm1)
    &=
    \frac{1}{2} \Big(
    \langle b_\perp^2 \rangle_{P^+}(\lambda=\pm1)
    +
    \langle b_\perp^2 \rangle_{P^+}(\lambda=0)
    \Big)
    \\
    \langle b_\perp^2 \rangle_{P^+}(m_s=0)
    &=
    \langle b_\perp^2 \rangle_{P^+}(\lambda=\pm1)
    \,,
  \end{align}
\end{subequations}
The average $P^+$ density between the three polarization states
(of massive hadrons)
is the same for helicity and transversely polarized states:
\begin{align}
  \overline{\rho}_{p^+}(b_\perp)
  &=
  \int \frac{\mathrm{d}^2\boldsymbol{\Delta}_\perp}{(2\pi)^2}
  \left\{
    \mathcal{G}_1(t)
    + \frac{t}{6 M^2} \left(
    \mathcal{G}_1(t)
    - \mathcal{G}_2(t)
    - \mathcal{G}_5(t)
    - \frac{1}{2} \mathcal{G}_6(t)
    \right)
    - \frac{t^2}{24 M^4} \mathcal{G}_2(t)
    \right\}
  e^{-i\boldsymbol{\Delta}_\perp\cdot\mathbf{b}_\perp}
  \,,
\end{align}
and likewise is the corresponding radius~\cite{Freese:2019bhb}:
\begin{align}
  \label{eqn:p+:radius:avg}
  \overline{\langle b_\perp^2 \rangle_{P^+}}
  &=
  4
  \frac{\mathrm{d}\mathcal{G}_1(t)}{\mathrm{d}t}
  \bigg|_{t=0}
  +
  \frac{2}{3M^2}
  \left(
  \mathcal{G}_1(0)
  - \mathcal{G}_2(0)
  - \mathcal{G}_5(0)
  - \frac{1}{2}\mathcal{G}_6(0)
  \right)
  \,.
\end{align}


\subsection{Angular momentum density}

The $J_z$ angular momentum density for helicity states
is given in Eq.~(\ref{eqn:J:hel}),
and for transversely polarized states is identically zero,
as already discussed in Sec.~\ref{sec:elements}.
As with the $P^+$ density, it may be helpful for numerical applications
to be able to take a single Hankel transform of $\mathcal{J}(t)$ itself.
Some straightforward algebra can be used to show that:
\begin{align}
  \rho_{J_z}^{(\lambda)}(b_\perp)
  =
  \frac{\lambda b_\perp}{2\pi}
  \mathscr{H}_1 \left[
    k \mathcal{J}(-k^2)
    \right](b_\perp)
  \,.
\end{align}
From this density,
the total angular momentum projected along the $z$ axis is:
\begin{align}
  \int \mathrm{d}^2 \mathbf{b}_\perp \,
  \rho_{J_z}^{(\lambda)}(b_\perp)
  =
  \lambda \mathcal{J}(0)
  =
  \frac{\lambda}{2} \mathcal{G}_5(0)
  \,.
\end{align}
Since this must be $\lambda$, we reproduce the finding
of Refs.\cite{Abidin:2008ku,Taneja:2011sy,Cosyn:2019aio}
that $\mathcal{G}_5(0)=2$.

For helicity $\lambda=\pm1$ states,
an angular momentum radius can be defined as:
\begin{align}
  \label{eqn:J:radius}
  \langle b_\perp^2 \rangle_{J}
  =
  \frac{1}{\lambda}
  \int \mathrm{d}^2\mathbf{b}_\perp \,
  \mathbf{b}_\perp^2
  \rho_{J_z}^{(\lambda)}(b_\perp)
  =
  8
  \frac{\mathrm{d}\mathcal{J}(t)}{\mathrm{d}t}
  \bigg|_{t=0}
  =
  4
  \frac{\mathrm{d}\mathcal{G}_5(t)}{\mathrm{d}t}
  \bigg|_{t=0}
  \,.
\end{align}


\subsection{The comoving stress tensor}

Following Refs.~\cite{Polyakov:2018zvc,Freese:2021czn},
the comoving stress tensor can most easily
be dealt with using the following auxiliary density
(which we call the Polyakov stress potential):
\begin{align}
  \label{eqn:Dtilde}
  \widetilde{\mathcal{D}}_\lambda(b_\perp)
  &=
  \frac{1}{4P^+}
  \int \frac{\mathrm{d}^2\boldsymbol{\Delta}_\perp}{(2\pi)^2}
  \mathcal{D}_{\lambda\lambda}(t)
  e^{-i\boldsymbol{\Delta}_\perp\cdot\mathbf{b}_\perp}
  \,,
\end{align}
for which
\begin{align}
  \label{eqn:Sij:Dtilde}
  S_\lambda^{ij}(\mathbf{b}_\perp)
  &=
  \Big(
  \boldsymbol{\nabla}_\perp^2 \delta_{ij}
  -
  \boldsymbol{\nabla}_\perp^i \boldsymbol{\nabla}_\perp^j
  \Big)
  \widetilde{\mathcal{D}}_\lambda(b_\perp)
  \,.
\end{align}

Analogously to spin-zero and spin-half helicity
states~\cite{Polyakov:2018zvc,Lorce:2018egm,Freese:2021czn},
the comoving stress tensor for spin-one helicity states can be decomposed
into an isotropic pressure $p(b_\perp)$ and shear stress
(or pressure anisotropy) function $s(b_\perp)$ as follows:
\begin{align}
  \label{eqn:Sij:decomp:hel}
  S_\lambda^{ij}(\mathbf{b}_\perp)
  =
  \delta^{ij} p^{(\lambda)}(b_\perp)
  +
  \left( \frac{b_\perp^i b_\perp^j}{b_\perp^2} - \frac{1}{2} \delta^{ij} \right)
  s^{(\lambda)}(b_\perp)
  \,.
\end{align}
This decomposition entails radial and tangential eigenpressures,
given by:
\begin{subequations}
  \label{eqn:prt:hel}
  \begin{align}
    \label{eqn:pr:hel}
    p_r^{(\lambda)}(b_\perp)
    &=
    p^{(\lambda)}(b_\perp)
    +
    \frac{s^{(\lambda)}(b_\perp)}{2}
    =
    \frac{1}{b_\perp}
    \frac{\mathrm{d} \widetilde{\mathcal{D}}_\lambda(b_\perp)}{\mathrm{d}b_\perp}
    \\
    p_t^{(\lambda)}(b_\perp)
    &=
    p^{(\lambda)}(b_\perp)
    -
    \frac{s^{(\lambda)}(b_\perp)}{2}
    =
    \frac{\mathrm{d}^2\widetilde{\mathcal{D}}_\lambda(b_\perp)}{\mathrm{d}b_\perp^2}
    \,.
  \end{align}
\end{subequations}
As with the $P^+$ density, it may be helpful for numerical applications
to obtain these quantities through higher-order Hankel transforms,
rather than through derivatives.
The isotropic pressure and shear stress can be shown to be:
\begin{subequations}
  \label{eqn:ps:iso}
  \begin{align}
    p^{(\lambda)}(b_\perp)
    &=
    -
    \frac{1}{8P^+}
    \frac{1}{2\pi}
    \mathscr{H}_0\left[
      k^2 \mathcal{D}_{\lambda\lambda}(-k^2)
      \right](b_\perp)
    \\
    s^{(\lambda)}(b_\perp)
    &=
    -
    \frac{1}{4P^+}
    \frac{1}{2\pi}
    \mathscr{H}_2\left[
      k^2 \mathcal{D}_{\lambda\lambda}(-k^2)
      \right](b_\perp)
    \,.
  \end{align}
\end{subequations}


\subsubsection{Transverse polarization}

For transversely polarized states, the structure of the comoving stress tensor
becomes significantly more complicated.
The Polyakov stress potential obtains modulations completely analogous
to those in the $P^+$ density;
we define:
\begin{subequations}
  \begin{align}
    \widetilde{D}_T^{\sin \phi }(b_\perp)
    &=
    \frac{1}{2\pi}
    \frac{1}{4P^+}
    \frac{1}{2M}
    \mathscr{H}_1\big[ k\mathcal{D}_T^{\sin\phi}(-k^2) \big](b_\perp)
    \\
    \widetilde{D}_T^{\cos 2\phi }(b_\perp)
    &=
    \frac{1}{2\pi}
    \frac{1}{4P^+}
    \frac{1}{4M^2}
    \mathscr{H}_2\big[ k^2\mathcal{D}_T^{\cos 2\phi}(-k^2) \big](b_\perp)
    \,,
  \end{align}
\end{subequations}
where the effective form factor modulations are as defined in Eq.~(\ref{eqn:AJD:mod}).
The Polyakov potentials for transversely polarized states are given by:
\begin{subequations}
  \label{eqn:Dtilde:trans}
  \begin{align}
    \widetilde{D}_T^{(\pm1)}(\mathbf{b}_\perp)
    &=
    \frac{
      \widetilde{D}_+(b_\perp) + \widetilde{D}_0(b_\perp)
    }{2}
    \pm
    \sin \phi \,
    \widetilde{D}_T^{\sin \phi }(b_\perp)
    +
    \frac{1}{2} \cos 2\phi \,
    \widetilde{D}_T^{\cos 2\phi }(b_\perp)
    \\
    \widetilde{D}_T^{(0)}(\mathbf{b}_\perp)
    &=
    \widetilde{D}_+(b_\perp)
    -
    \cos 2\phi \,
    \widetilde{D}_T^{\cos 2\phi }(b_\perp)
    \,,
  \end{align}
\end{subequations}
where we have used
$\widetilde{D}_+(b_\perp) = \widetilde{D}_-(b_\perp)$
to make the formulas slightly shorter.
The comoving stress tensor is then given by:
\begin{align}
  \label{eqn:Sij:Dtilde:trans}
  S_T^{ij}(\mathbf{b}_\perp,m_s)
  &=
  \Big(
  \boldsymbol{\nabla}_\perp^2 \delta_{ij}
  -
  \boldsymbol{\nabla}_\perp^i \boldsymbol{\nabla}_\perp^j
  \Big)
  \widetilde{\mathcal{D}}_T^{(m_s)}(\mathbf{b}_\perp)
  \,.
\end{align}
This stress tensor no longer has the simple decomposition of
Eq.~(\ref{eqn:Sij:decomp:hel});
it contains a new tensor structure, and the functions multiplying
each structure now contain azimuthal modulations:
\begin{align}
  \label{eqn:Sij:trans}
  S^{ij}_T(\mathbf{b}_\perp, m_s)
  =
  \delta^{ij} p_T^{(m_s)}(\mathbf{b}_\perp)
  +
  \left( \hat{b}^i \hat{b}^j - \frac{1}{2} \delta^{ij} \right)
  s_T^{(m_s)}(\mathbf{b}_\perp)
  +
  \Big( \hat{b}^i \hat{\phi}^j + \hat{\phi}^i \hat{b}^j \Big)
  v_T^{(m_s)}(\mathbf{b}_\perp)
  \,.
\end{align}
Here, $\hat{b}$ and $\hat{\phi}$ are unit vectors in the radial
and counterclockwise tangential directions, respectively.
Note that
each of the tensor structures except for the $\delta^{ij}$
accompanying $p_T^{(m_s)}(\mathbf{b}_\perp)$ is traceless,
so $p_T^{(m_s)}(\mathbf{b}_\perp)$ can be understood as the
isotropic pressure.

To obtain the functions $p_T$, $s_T$, and $v_T$,
one can contract the comoving stress tensor with
multiples of the associated tensors:
\begin{subequations}
  \begin{align}
    p_T^{(m_s)}(\mathbf{b}_\perp)
    &=
    \frac{1}{2} \delta_{ij}
    S^{ij}_T(\mathbf{b}_\perp, m_s)
    \\
    s_T^{(m_s)}(\mathbf{b}_\perp)
    &=
    \left( \hat{b}^i \hat{b}^j - \frac{1}{2} \delta^{ij} \right)
    S^{ij}_T(\mathbf{b}_\perp, m_s)
    \\
    v_T^{(m_s)}(\mathbf{b}_\perp)
    &=
    \frac{1}{2}
    \Big( \hat{b}^i \hat{\phi}^j + \hat{\phi}^i \hat{b}^j \Big)
    S^{ij}_T(\mathbf{b}_\perp, m_s)
    \,.
  \end{align}
\end{subequations}
With some straightforward but tedious algebra,
combining these equations with Eq.~(\ref{eqn:Sij:Dtilde:trans}) yields:
\begin{subequations}
  \begin{align}
    p_T^{(m_s)}(\mathbf{b}_\perp)
    &=
    \frac{1}{2}
    \left\{
      \frac{\partial^2}{\partial b_\perp^2}
      +
      \frac{1}{b_\perp}
      \frac{\partial}{\partial b_\perp}
      +
      \frac{1}{b_\perp^2}
      \frac{\partial^2}{\partial \phi^2}
      \right\}
    \widetilde{\mathcal{D}}_T^{(m_s)}(\mathbf{b}_\perp)
    \\
    s_T^{(m_s)}(\mathbf{b}_\perp)
    &=
    \left\{
      -
      \frac{\partial^2}{\partial b_\perp^2}
      +
      \frac{1}{b_\perp}
      \frac{\partial}{\partial b_\perp}
      +
      \frac{1}{b_\perp^2}
      \frac{\partial^2}{\partial \phi^2}
      \right\}
    \widetilde{\mathcal{D}}_T^{(m_s)}(\mathbf{b}_\perp)
    \\
    v_T^{(m_s)}(\mathbf{b}_\perp)
    &=
    \left\{
      -
      \frac{1}{b_\perp}
      \frac{\partial}{\partial b_\perp}
      +
      \frac{1}{b_\perp^2}
      \right\}
    \frac{\partial}{\partial \phi}
    \widetilde{\mathcal{D}}_T^{(m_s)}(\mathbf{b}_\perp)
    \,.
  \end{align}
\end{subequations}

It will be helpful for numerical applications to have expressions
for the functions $p_T$, $s_T$, and $v_T$ in terms of Hankel transforms
rather than coordinate derivatives.
Some algebra and identities for Bessel functions can be used to accomplish this.
We spare the reader the details of the derivation,
stating only the results.
For specific polarization states, these functions are given by:
\begin{subequations}
  \label{eqn:ps:trans}
  \begin{align}
    p_T^{(\pm 1)}(\mathbf{b}_\perp)
    &=
    \frac{p^{(+)}(b_\perp)+p^{(0)}(b_\perp)}{2}
    \pm
    \sin\phi \,
    p_T^{\sin \phi}(b_\perp)
    +
    \frac{1}{2}
    \cos 2\phi \,
    p_T^{\cos 2\phi}(b_\perp)
    \\
    s_T^{(\pm 1)}(\mathbf{b}_\perp)
    &=
    \frac{s^{(+)}(b_\perp)+s^{(0)}(b_\perp)}{2}
    \pm
    \sin\phi \,
    s_T^{\sin \phi}(b_\perp)
    +
    \frac{1}{2}
    \cos 2\phi \,
    s_T^{\cos 2\phi}(b_\perp)
    \\
    v_T^{(\pm 1)}(\mathbf{b}_\perp)
    &=
    \cos\phi \,
    v_T^{\cos \phi}(b_\perp)
    +
    \frac{1}{2}
    \sin 2\phi \,
    v_T^{\sin 2\phi}(b_\perp)
    \,,
  \end{align}
  and for the $m_s=0$ state are:
  \begin{align}
    p_T^{(0)}(\mathbf{b}_\perp)
    &=
    p^{(+)}(b_\perp)
    -
    \cos 2\phi \,
    p_T^{\cos 2\phi}(b_\perp)
    \\
    s_T^{(0)}(\mathbf{b}_\perp)
    &=
    s^{(+)}(b_\perp)
    -
    \cos 2\phi \,
    s_T^{\cos 2\phi}(b_\perp)
    \\
    v_T^{(0)}(\mathbf{b}_\perp)
    &=
    -
    \sin 2\phi \,
    v_T^{\sin 2\phi}(b_\perp)
    \,.
  \end{align}
\end{subequations}
The $\phi$ modulations in these functions are given by:
\begin{subequations}
  \label{eqn:ps:mod}
  \begin{align}
    p_T^{\sin \phi}(b_\perp)
    &=
    \frac{1}{2\pi}
    \frac{1}{4P^+}
    \frac{1}{2M}
    \left\{
      -
      \frac{1}{2}
      \mathscr{H}_1 \left[
        k^3 \mathcal{D}_T^{\sin\phi}(-k^2)
        \right](b_\perp)
      \right\}
    \\
    s_T^{\sin \phi}(b_\perp)
    &=
    \frac{1}{2\pi}
    \frac{1}{4P^+}
    \frac{1}{2M}
    \left\{
      \mathscr{H}_1 \left[
        k^3 \mathcal{D}_T^{\sin\phi}(-k^2)
        \right](b_\perp)
      -
      \frac{2}{b_\perp}
      \mathscr{H}_2 \left[
        k^2 \mathcal{D}_T^{\sin\phi}(-k^2)
        \right](b_\perp)
      \right\}
    \\
    v_T^{\cos \phi}(b_\perp)
    &=
    \frac{1}{2\pi}
    \frac{1}{4P^+}
    \frac{1}{2M}
    \left\{
      \frac{1}{b_\perp}
      \mathscr{H}_2 \left[
        k^2 \mathcal{D}_T^{\sin\phi}(-k^2)
        \right](b_\perp)
      \right\}
    \,,
  \end{align}
  and the $2\phi$ modulations by:
  \begin{align}
    p_T^{\cos 2\phi}(b_\perp)
    &=
    \frac{1}{2\pi}
    \frac{1}{4P^+}
    \frac{1}{4M^2}
    \left\{
      -
      \frac{1}{2}
      \mathscr{H}_2 \left[
        k^4 \mathcal{D}_T^{\cos 2\phi}(-k^2)
        \right](b_\perp)
      \right\}
    \\
    s_T^{\cos 2\phi}(b_\perp)
    &=
    \frac{1}{2\pi}
    \frac{1}{4P^+}
    \frac{1}{4M^2}
    \left\{
      -
      \frac{1}{2}
      \mathscr{H}_0 \left[
        k^4 \mathcal{D}_T^{\cos 2\phi}(-k^2)
        \right](b_\perp)
      -
      \frac{1}{2}
      \mathscr{H}_4 \left[
        k^4 \mathcal{D}_T^{\cos 2\phi}(-k^2)
        \right](b_\perp)
      \right\}
    \\
    v_T^{\sin 2\phi}(b_\perp)
    &=
    \frac{1}{2\pi}
    \frac{1}{4P^+}
    \frac{1}{4M^2}
    \left\{
      \frac{1}{4}
      \mathscr{H}_0 \left[
        k^4 \mathcal{D}_T^{\cos 2\phi}(-k^2)
        \right](b_\perp)
      -
      \frac{1}{4}
      \mathscr{H}_4 \left[
        k^4 \mathcal{D}_T^{\cos 2\phi}(-k^2)
        \right](b_\perp)
      \right\}
    \,.
  \end{align}
\end{subequations}

For transversely polarized states, the eigenpressures will no longer
be radial and tangential.
The eigenpressures are instead given by\footnote{
  A capital $P$ is used to signify transverse eigenpressures to assist
  visually distingiushing them from other auxilliary functions
  such as $p_T^{(m_s)}$.
}:
\begin{subequations}
  \label{eqn:eigenpressure}
  \begin{align}
    P_{T,\pm}^{(m_s)}(\mathbf{b}_\perp)
    =
    p_T^{(m_s)}(\mathbf{b}_\perp)
    \pm
    \sqrt{
      \frac{1}{4} \big(s_T^{(m_s)}(\mathbf{b}_\perp)\big)^2
      +
      \big(v_T^{(m_s)}(\mathbf{b}_\perp)\big)^2
    }
    \,.
  \end{align}
  These eigenpressures are normal stresses along $\phi$-dependent
  unit vectors $\hat{e}_\pm$, whose angles with respect to the spin quantization
  axis $\mathbf{s}_\perp = \hat{x}$ are given by:
  \begin{align}
    \theta_\pm^{(m_s)}(\mathbf{b}_\perp)
    &=
    \phi
    +
    \frac{1}{2}
    \tan^{-1}\left(
    \frac{ 2 v_T^{(m_s)}(\mathbf{b}_\perp) }{ s_T^{(m_s)}(\mathbf{b}_\perp) }
    \right)
    +
    \Theta\Big(\pm s_T^{(m_s)}(\mathbf{b}_\perp)\Big)
    \frac{\pi}{2}
    \,,
  \end{align}
  where $\Theta(x)$ is the Heaviside step function.
  The unit eigenvectors are then written:
  \begin{align}
    \hat{e}_\pm^{(m_s)}(\mathbf{b}_\perp)
    &=
    \cos\big(\theta_\pm^{(m_s)}(\mathbf{b}_\perp)\big)
    \, \hat{x}
    +
    \sin\big(\theta_\pm^{(m_s)}(\mathbf{b}_\perp)\big)
    \, \hat{y}
    \,.
  \end{align}
\end{subequations}
It is also possible to categorize the eigenpressures in an alternative way:
\begin{subequations}
  \label{eqn:eigenpressure:alt}
  \begin{align}
    \bar{P}_{T,r}^{(m_s)}(\mathbf{b}_\perp)
    &=
    p_T^{(m_s)}(\mathbf{b}_\perp)
    +
    \mathrm{sign}\big( s_T^{(m_s)}(\mathbf{b}_\perp)\big)
    \sqrt{
      \frac{1}{4} \big(s_T^{(m_s)}(\mathbf{b}_\perp)\big)^2
      +
      \big(v_T^{(m_s)}(\mathbf{b}_\perp)\big)^2
    }
    \\
    \bar{P}_{T,t}^{(m_s)}(\mathbf{b}_\perp)
    &=
    p_T^{(m_s)}(\mathbf{b}_\perp)
    -
    \mathrm{sign}\big( s_T^{(m_s)}(\mathbf{b}_\perp)\big)
    \sqrt{
      \frac{1}{4} \big(s_T^{(m_s)}(\mathbf{b}_\perp)\big)^2
      +
      \big(v_T^{(m_s)}(\mathbf{b}_\perp)\big)^2
    }
    \,,
  \end{align}
  whose angles with respect to the spin quantization axis are:
  \begin{align}
    \bar\theta_r^{(m_s)}(\mathbf{b}_\perp)
    &=
    \phi
    +
    \frac{1}{2}
    \tan^{-1}\left(
    \frac{ 2 v_T^{(m_s)}(\mathbf{b}_\perp) }{ s_T^{(m_s)}(\mathbf{b}_\perp) }
    \right)
    \\
    \bar\theta_t^{(m_s)}(\mathbf{b}_\perp)
    &=
    \phi
    +
    \frac{1}{2}
    \tan^{-1}\left(
    \frac{ 2 v_T^{(m_s)}(\mathbf{b}_\perp) }{ s_T^{(m_s)}(\mathbf{b}_\perp) }
    \right)
    +
    \frac{\pi}{2}
    \,.
  \end{align}
\end{subequations}
At every $\mathbf{b}_\perp$, these of course furnish the same pair of
eigenvectors and eigenvalues as Eq.~(\ref{eqn:eigenpressure});
the difference lies in how the pairs are sorted into
$\mathbf{b}_\perp$-dependent functions.
The eigenvalue/eigenvector pairs in Eq.~(\ref{eqn:eigenpressure:alt})
in particular reduce to the familiar radial and tangential eigenpressures
in the helicity case (where $v_T=0$).
However, there is benefit to using Eq.~(\ref{eqn:eigenpressure})
instead of Eq.~(\ref{eqn:eigenpressure:alt}) for transversely polarized states:
namely, that when $v_T\neq0$, only the former are continuous across $s_T=0$.
This can be seen both in the square root function in the pressure functions themselves,
and in how the step function in the angle functions compensates the $\frac{\pi}{2}$ discontinuity
between $\frac{1}{2}\tan^{-1}(\infty)$ and between $\frac{1}{2}\tan^{-1}(-\infty)$.


\subsubsection{Mechanical radius}

It has been hypothesized~\cite{Perevalova:2016dln,Polyakov:2018zvc,Lorce:2018egm,Freese:2021czn}
that the radial pressure is a positive-definite quantity for stable systems,
and can thus be used to define a positive-definite ``mechanical radius''
that gives an estimate of a hadron's size:
\begin{align}
  \label{eqn:r:mech}
  \langle b_\perp^2 \rangle_{\mathrm{mech}}
  =
  \frac{
    \int \mathrm{d^2}\mathbf{b}_\perp \,
    \mathbf{b}_\perp^2
    p_r(\mathbf{b_\perp})
  }{
    \int \mathrm{d^2}\mathbf{b}_\perp \,
    p_r(\mathbf{b_\perp})
  }
  \,.
\end{align}
For transversely polarized states the radial pressure is not an eigenpressure,
but it is nevertheless a normal stress along the $\hat{b}$ direction,
and is given by:
\begin{align}
  p_{Tr}^{(m_s)}(\mathbf{b}_\perp)
  =
  \frac{1}{b_\perp}
  \frac{
    \partial \widetilde{\mathcal{D}}_T^{(m_s)}(\mathbf{b}_\perp)
  }{\partial b_\perp}
  +
  \frac{
    \partial^2 \widetilde{\mathcal{D}}_T^{(m_s)}(\mathbf{b}_\perp)
  }{\partial \phi^2}
  \,.
\end{align}
In both the numerator and denominator,
the integrals over the azimuthal modulations become zero.
Thus, for either helicity or transversely polarized states,
the numerator becomes, via integration by parts:
\begin{align}
  \int \mathrm{d^2}\mathbf{b}_\perp \,
  \mathbf{b}_\perp^2
  p_r(\mathbf{b_\perp})
  =
  -
  \frac{1}{2P^+} \mathcal{D}(0)
  \,.
\end{align}
The denominator, with a little integration calculus, can be shown to be:
\begin{align}
  \label{eqn:mech:denominator}
  \int \mathrm{d^2}\mathbf{b}_\perp \,
  p_r(\mathbf{b_\perp})
  =
  -
  \frac{1}{16\pi P^+}
  \int_{-\infty}^0 \mathrm{d}t \,
  \int_0^{2\pi} \mathrm{d}\phi \,
  \mathcal{D}(\boldsymbol{\Delta}_\perp)
  \,,
\end{align}
where the modulations again integrate to zero.
The mechanical radius is thus given by:
\begin{align}
  \label{eqn:mechrad:D}
  \langle b_\perp^2 \rangle_{\mathrm{mech}}
  =
  \frac{
    4 \mathcal{D}(0)
  }{
    \int_{-\infty}^0 \mathrm{d}t \,
    \mathcal{D}(t)
  }
  \Bigg|_{\sin\phi=0,\,\cos 2\phi=0}
  \,.
\end{align}
For specific helicity states of massive hadrons, we have:
\begin{subequations}
  \begin{align}
    \langle b_\perp^2 \rangle_{\mathrm{mech}}(\lambda=\pm1)
    &=
    \frac{
      4 \big( \mathcal{G}_3(0) - \mathcal{G}_6(0) \big)
    }{
      \int_{-\infty}^0 \mathrm{d}t \,
      \big( \mathcal{G}_3(t) - \mathcal{G}_6(t) \big)
    }
    \\
    \langle b_\perp^2 \rangle_{\mathrm{mech}}(\lambda=0)
    &=
    \frac{
      4 \mathcal{G}_3(0)
    }{
      \int_{-\infty}^0 \mathrm{d}t \,
      \mathcal{G}_3(t)
    }
    \,,
  \end{align}
  while for specific transverse polarization states, we have:
  \begin{align}
    \langle b_\perp^2 \rangle_{\mathrm{mech},T}(m_s=\pm1)
    &=
    \frac{
      4 \big(
      \mathcal{G}_3(0) - \frac{1}{2} \mathcal{G}_6(0)
      \big)
    }{
      \int_{-\infty}^0 \mathrm{d}t \,
      \big( \mathcal{G}_3(t) - \frac{1}{2}\mathcal{G}_6(t) \big)
    }
    \\
    \langle b_\perp^2 \rangle_{\mathrm{mech},T}(m_s=0)
    &=
    \frac{
      4 \big( \mathcal{G}_3(0) - \mathcal{G}_6(0) \big)
    }{
      \int_{-\infty}^0 \mathrm{d}t \,
      \big( \mathcal{G}_3(t) - \mathcal{G}_6(t) \big)
    }
    \,.
  \end{align}
\end{subequations}
For the unpolarized state, the numerator and denominator need to be
averaged separately.
The unpolarized mechanical radius of a massive hadron is given by:
\begin{align}
  \overline{\langle b_\perp^2 \rangle_{\mathrm{mech}}}
  =
  \frac{
    4 \big(
    \mathcal{G}_3(0) - \frac{2}{3} \mathcal{G}_6(0)
    \big)
  }{
    \int_{-\infty}^0 \mathrm{d}t \,
    \big( \mathcal{G}_3(t) - \frac{2}{3}\mathcal{G}_6(t) \big)
  }
  \,.
\end{align}


\section{Numerical illustration}
\label{sec:example}

As a simple numerical illustration, we present light front densities for the
deuteron in a light cone convolution model~\cite{Cano:2003ju,Cosyn:2017fbo,Cosyn:2018rdm,Cosyn:2020kwu}.
The model provides a description of deuteron structure in terms of
on-shell nucleons, which allows for on-shell gravitational form factors
to be used for the nucleon,
according to the standard formulas (e.g.\ Eq.~(6) of Ref.~\cite{Polyakov:2018zvc}).

A potential downside of the light cone model is that it breaks manifest
Lorentz covariance by truncating the Fock state at a two-nucleon state---a
truncation that is invariant under the kinematic subgroup,
but not under dynamical transformations.
The form factor and helicity amplitude
breakdowns in Eqs.~(\ref{eqn:emt:massive}) and (\ref{eqn:emt:spin1})
are a consequence of Lorentz covariance,
and accordingly, the
helicity amplitudes
calculated in this model through
different components of the EMT may be inconsistent.
(Compare to Refs.~\cite{Cano:2003ju,Cosyn:2018rdm}, where polynomiality breaks down for
generalized parton distributions of the deuteron,
which makes extraction of the GFFs ambiguous.)
Additionally, the components $T^{+i}$ and $T^{ij}$ are ``bad'' components~\cite{Melosh:1974cu,Leutwyler:1977vy},
in the sense that they mix Fock states with different numbers of particles,
and the truncation of the deuteron Fock state at two nucleons accordingly
drops potentially relevant physics.

Despite this potential shortcoming, we adopt the model in question,
largely due to the lack of alternatives with the desirable covariance property.
Moreover, this section is primarily meant to illustrate the general formalism
developed above---a purpose for which the model is perfectly adequate.
To deal with the issue of inconsistent
helicity amplitudes,
we consider specifically components of the EMT that give expected behavior
of the GFFs at $t=0$, namely that the $t=0$ results for all helicity transitions
are zero (e.g., $\mathcal{J}_{+0}(0) = 0$),
and that $\mathcal{D}_{\lambda\lambda'}(0)$ is finite for all $\lambda$ and $\lambda'$.

We calculate in a frame where $\mathbf{P}_\perp=0$.
We have $\Delta^+=0$ by construction, and it also follows that $\Delta^-=0$.
Without loss of generality, we can consider $\Delta^x=\sqrt{-t}$ and $\Delta^y=0$.
In the convolution the momentum of the ``active'' nucleon enters the matrix element of the EMT.
This has the same $\Delta^+$ and $\Delta_\perp$ as for the deuteron,
whereas $\Delta^-$ does not enter into the relevant matrix elements.

We find the following EMT matrix elements to provide GFFs with the
required $t=0$ behavior:
\begin{subequations}
  \label{eqn:conv:deuteron}
  \begin{align}
    \frac{1}{2(P^+)^2}\langle p'\lambda'|T^{++}|p\lambda\rangle
    &=
    \mathcal{A}_{\lambda'\lambda}(t)
    \,,
    \\
    -\frac{1}{P^+\sqrt{-t}} \,\langle p'\lambda'|T^{+R}|p\lambda\rangle
    &=
    \mathcal{J}_{\lambda'\lambda}(t)
    \,,
    \\
    -2 \langle p'\lambda'|T^{RR}|p\lambda\rangle
    &=
    t\mathcal{D}_{\lambda'\lambda}
    \,,
  \end{align}
\end{subequations}
where the $R$ and $L$ components are defined via:
\begin{subequations}
  \begin{align}
    a^R &= a^x + ia^y
    \,,
    \\
    a^L &= a^x - ia^y
    \,.
  \end{align}
\end{subequations}
These allow us to calculate the necessary
helicity amplitudes
directly.
Specifically, off-diagonal matrix elements contribute to
the $\sin\phi$ (one unit helicity difference)
and $\cos2\phi$ (two units) modulations for the transversely polarized states.

Because of Lorentz covariance violations by the convolution model,
several symmetry relations laid out in Appendix~\ref{app:elements} are violated
by applying Eq.~(\ref{eqn:conv:deuteron}) to the model.
For instance, we find $\mathcal{J}_{+0}(t) \neq - \mathcal{J}_{0+}^*(t)$.
In this case specifically,
we find $\mathcal{J}_{+0}(0)=0$ but $\mathcal{J}_{0+}(0)\neq0$.
Since physically this
helicity amplitude
should vanish at $t=0$,
we use Eq.~(\ref{eqn:conv:deuteron}) to calculate $\mathcal{J}_{+0}(t)$ specifically,
and then set $\mathcal{J}_{0+}(t) = -\mathcal{J}_{+0}^*(t)$.
(If we use $T^{+L}$ rather than $T^{+R}$ to calculate these same
helicity amplitudes,
their behavior is actually reversed.
This behavior reversal is an inevitable consequence of stricter symmetry
properties than Lorentz covariance,
namely hermiticity and parity invariance.)
In all cases where the relations in Appendix~\ref{app:elements} are violated,
we restore the relations by fiat and use
Eq.~(\ref{eqn:conv:deuteron}) to calculate the specific
helicity amplitude
with the required $t=0$ behavior.

To proceed, we also need the following matrix elements for the nucleon light front EMT,
obtained by evaluating Eq.~(27) of Ref.~\cite{Freese:2021czn}.
For $T^{++}$ matrix elements:
\begin{subequations}
  \begin{align}
    \frac{1}{2(P^+)^2}\langle p'_N\lambda|T^{++}|p_N\lambda\rangle &= \left(\frac{\alpha_N}{2}\right)^2 A(t)\,,\\
    \frac{1}{2(P^+)^2}\langle p'_N \,-|T^{++}|p_N \, +\rangle &=
    -\left(\frac{\alpha_N}{2}\right)^2 \frac{\sqrt{-t}}{2M} \left[ A(t) - 2J(t) \right]\,,\\
    \frac{1}{2(P^+)^2}\langle p'_N \,+|T^{++}|p_N \, -\rangle &=
    \left(\frac{\alpha_N}{2}\right)^2 \frac{\sqrt{-t}}{2M} \left[ A(t) - 2J(t) \right]\,,
  \end{align}
\end{subequations}
for $T^{+R}$ matrix elements:
\begin{subequations}
  \begin{align}
    -\frac{1}{P^+\sqrt{-t}}\langle p'_N\lambda|T^{+R}|p_N\lambda\rangle &=
    \frac{\alpha_N}{2}\left(  -2 \frac{P_N^R}{\sqrt{-t}}A(t)+\lambda J(t) \right)\,,\\
    -\frac{1}{P^+\sqrt{-t}}\langle p'_N \,-|T^{+R}|p_N \, +\rangle &=
    \frac{\alpha_N}{2} \frac{P_N^R}{M} \left[  A(t) -2 J(t) \right]\,,\\
    -\frac{1}{P^+\sqrt{-t}}\langle p'_N \,+|T^{+R}|p_N \, -\rangle &=
    -\frac{\alpha_N}{2} \frac{P_N^R}{M} \left[  A(t) + \left(\frac{P_N^L}{P_N^R}-1 \right) J(t)  \right]\,,
  \end{align}
\end{subequations}
and for $T^{RR}$ matrix elements:
\begin{subequations}
  \begin{align}
    2\langle p'_N\lambda|T^{RR}|p_N\lambda\rangle &=
    4(P_N^R)^2 A(t) - tD(t) - 2\lambda \sqrt{-t}P_N^R J(t)\,,\\
    2\langle p'_N \,-|T^{RR}|p_N \, +\rangle &=
    -\frac{\sqrt{-t}}{2M}\left[4(P_N^R)^2 A(t) - tD(t)\right] + 2\frac{\sqrt{-t}}{M} (P_N^R)^2 J(t)\,,\\
    2\langle p'_N \,+|T^{RR}|p_N \, -\rangle &=
    \frac{\sqrt{-t}}{2M}\left[4(P_N^R)^2 A(t) - tD(t)\right] + 2\frac{\sqrt{-t}}{M} P_N^RP_N^L J(t)\,.
  \end{align}
\end{subequations}
Here, $\alpha_N$ is related to the light front momentum fraction of the active nucleon:
\begin{equation}
    \alpha_N = \frac{2p_N^+}{p^+} = \frac{2p_N^{'+}}{p^+}\,.
\end{equation}
For the nucleon form factors, we use simple multipole parametrizations,
motivated by the investigations of Ref.~\cite{Masjuan:2012sk}
(see Sec.~V.C thereof in particular):
\begin{subequations}
  \begin{align}
    A(t)
    &=
    2 J(t)
    =
    \frac{1}{
      \left(1-t/m_{f_2(1270)}^2\right)
      \left(1-t/m_{f_2(1430)}^2\right)
    }
    \,,
    \\
    D(t)
    &=
    \frac{
      D(0)
    }{
      \left(1-t/m_{f_2(1270)}^2\right)
      \left(1-t/m_{f_2(1430)}^2\right)
      \left(1-t/m_{\sigma(800)}^2\right)
    }
    \,,
  \end{align}
\end{subequations}
with $D(0) = -2$,
motivated by lattice QCD findings~\cite{Pefkou:2021fni}.

\begin{figure}
  \includegraphics[width=0.32\textwidth]{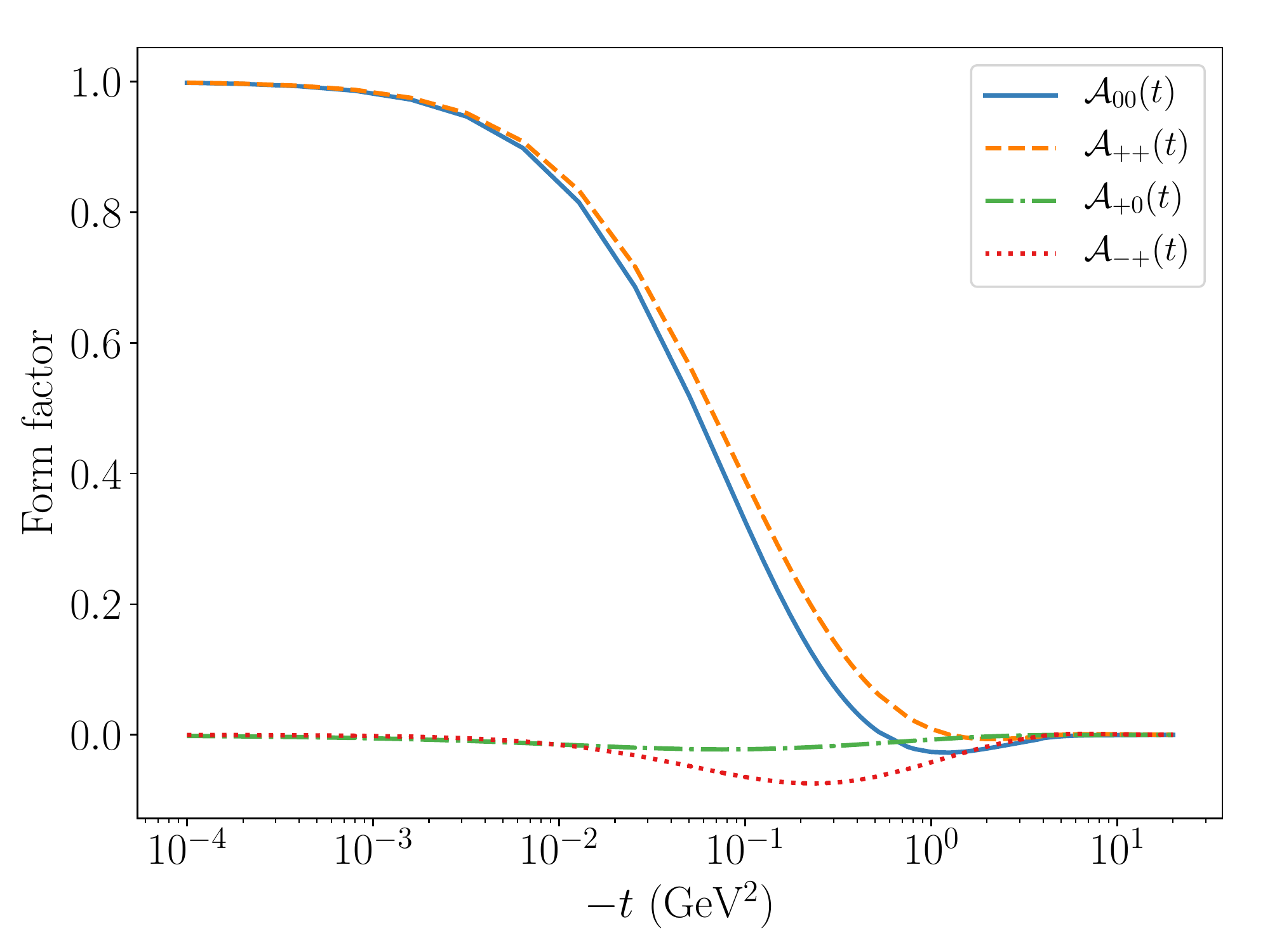}
  \includegraphics[width=0.32\textwidth]{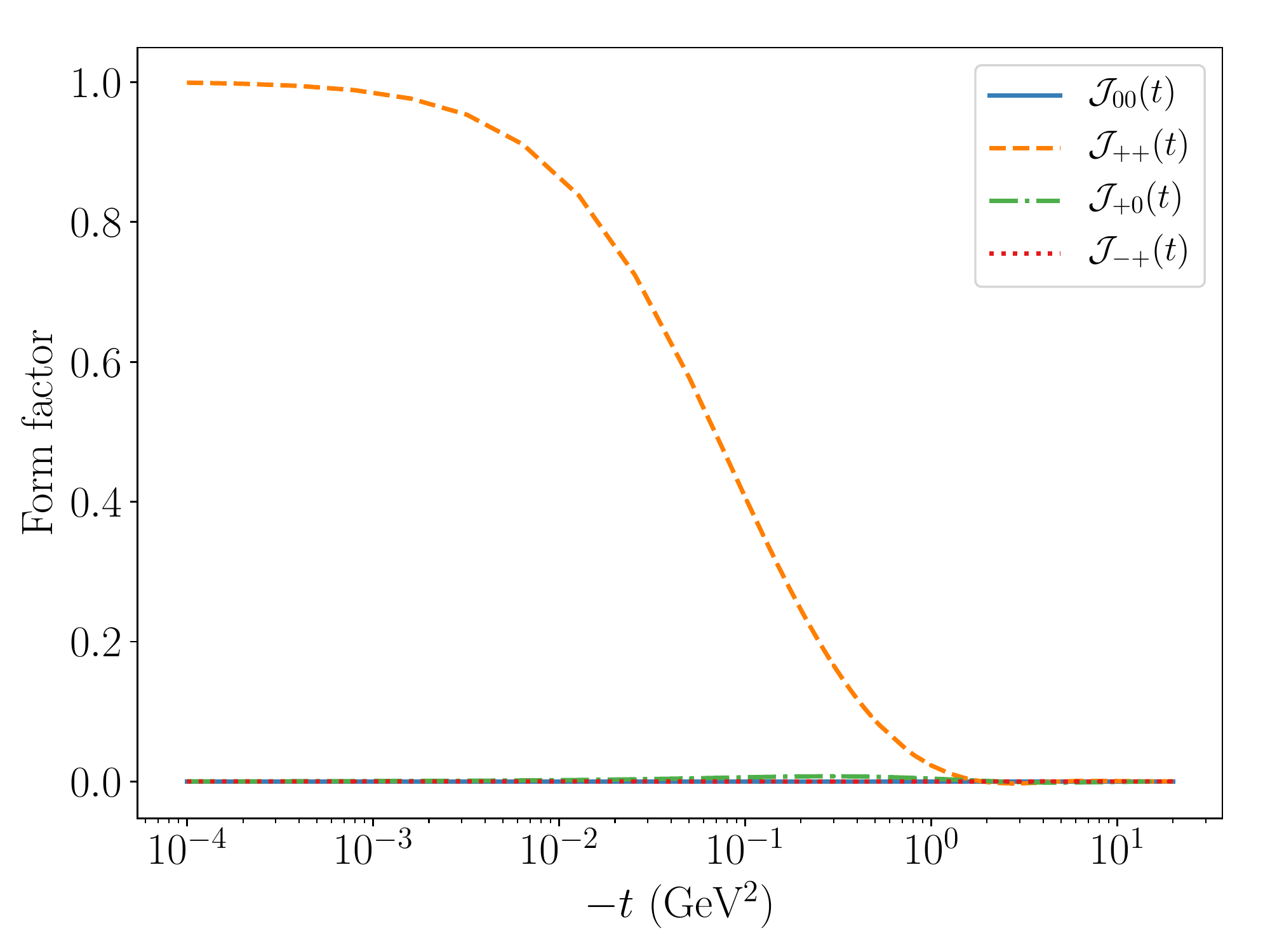}
  \includegraphics[width=0.32\textwidth]{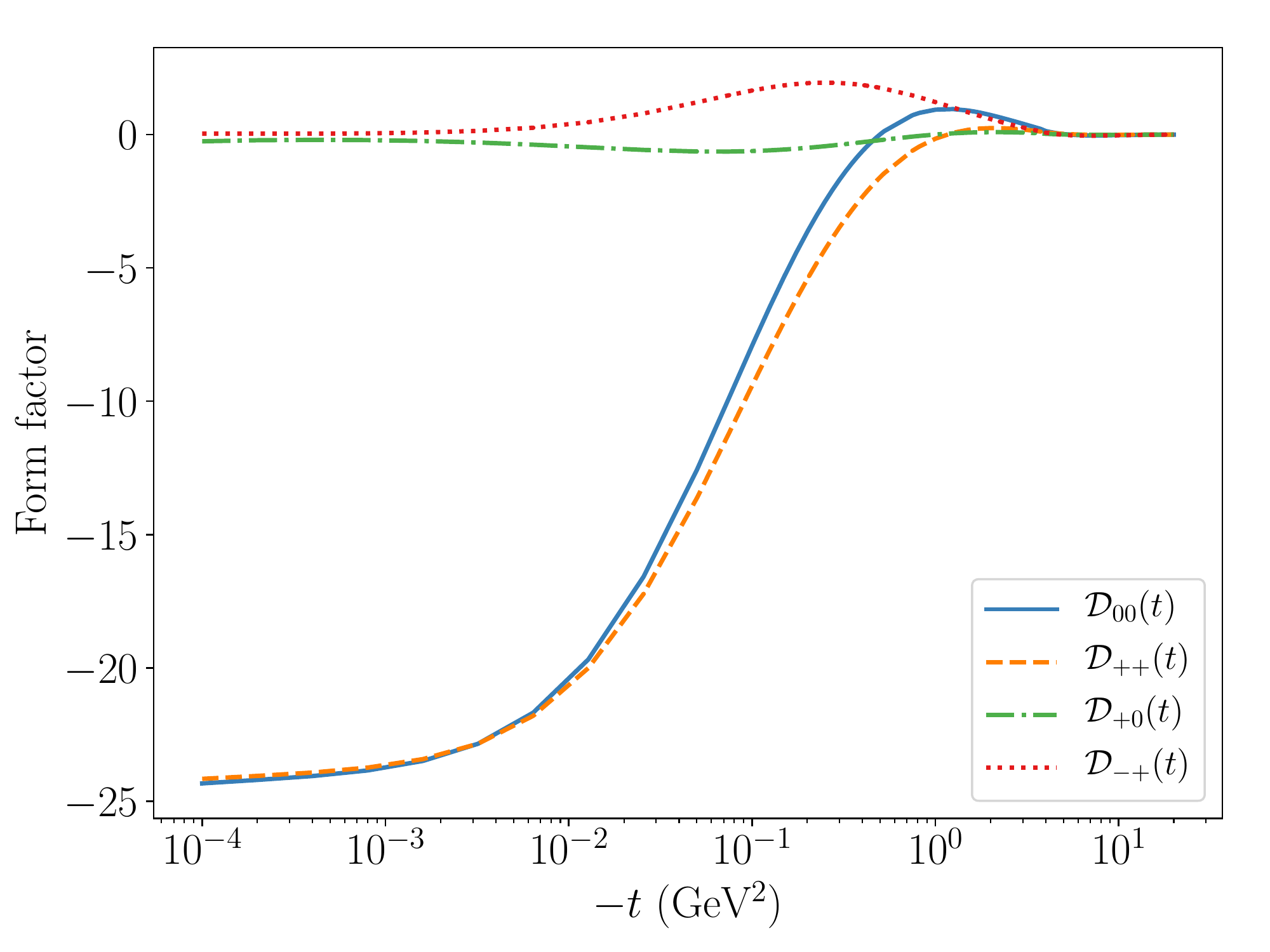}
  \caption{
    The helicity amplitudes $\mathcal{A}(t)$, $\mathcal{J}(t)$
    and $\mathcal{D}(t)$ of the deuteron for various helicity
    combinations.
    Those not explicitly given in the plots are determined from these
    using the relations in Appendix~\ref{app:elements}.
  }
  \label{fig:conv:gff}
\end{figure}

The
helicity amplitudes
in this model are presented
in Fig.~\ref{fig:conv:gff}.
From these, a variety of light front densities can be obtained.
We present a selected sample of these densities,
in order to not take up too much space.
In particular, $p^+$ densities can be calculated using
Eqs.~(\ref{eqn:p+:hel}) and (\ref{eqn:hankel:A}),
and the pressure distributions using
Eqs.~(\ref{eqn:ps:iso}), (\ref{eqn:ps:trans}), (\ref{eqn:ps:mod})
and (\ref{eqn:eigenpressure}).

\begin{figure}
  \includegraphics[width=0.49\textwidth]{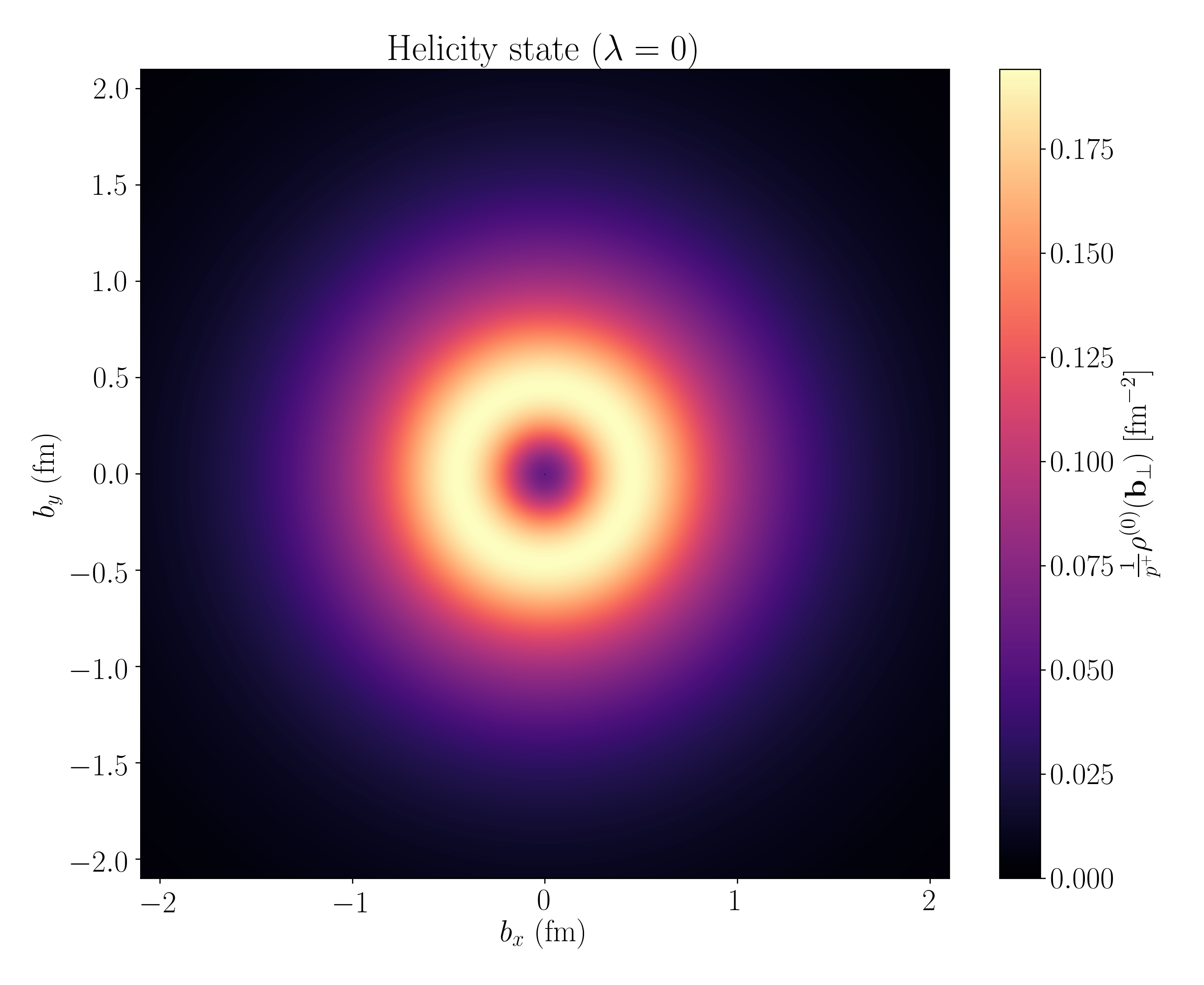}
  \includegraphics[width=0.49\textwidth]{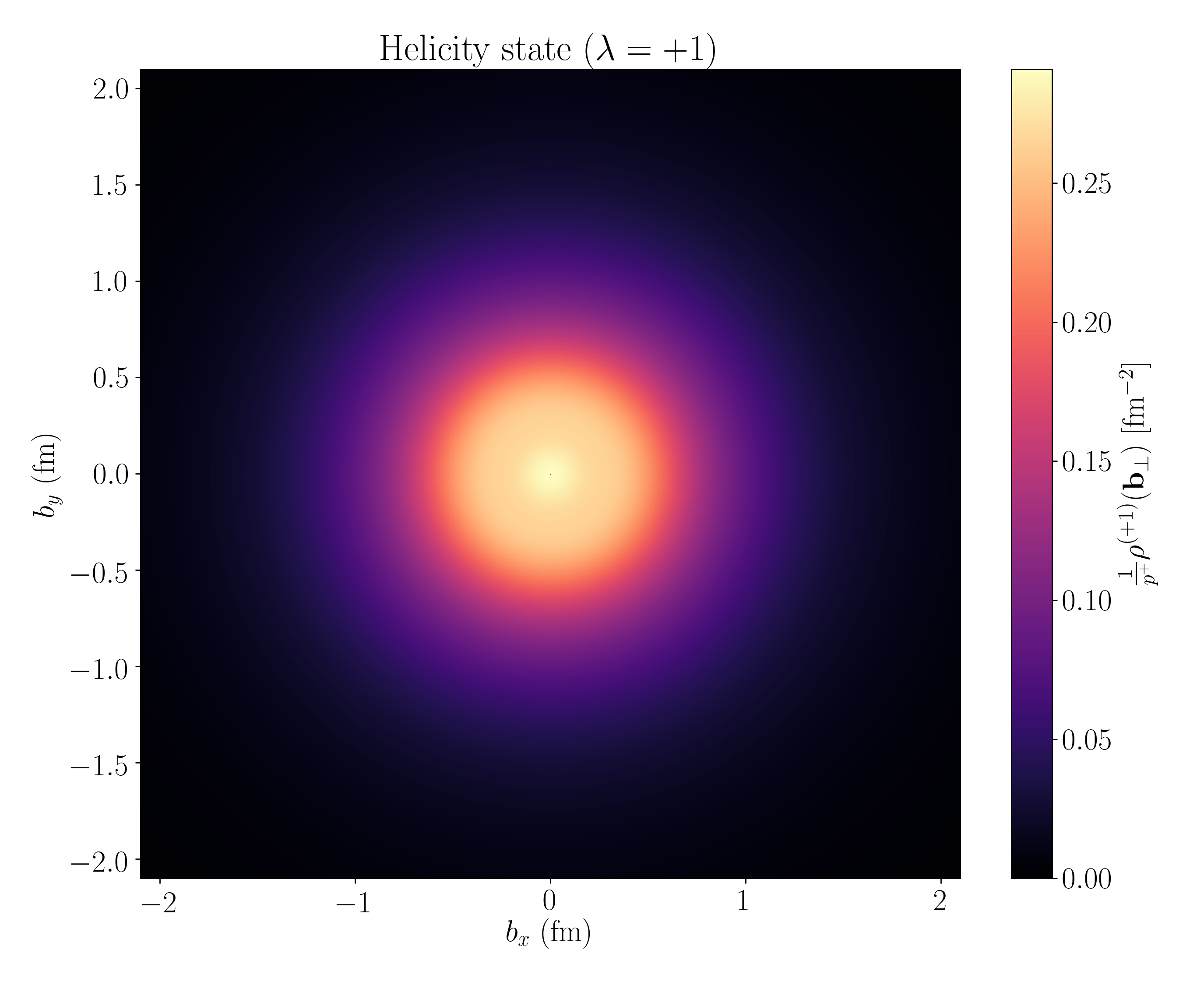}
  \includegraphics[width=0.49\textwidth]{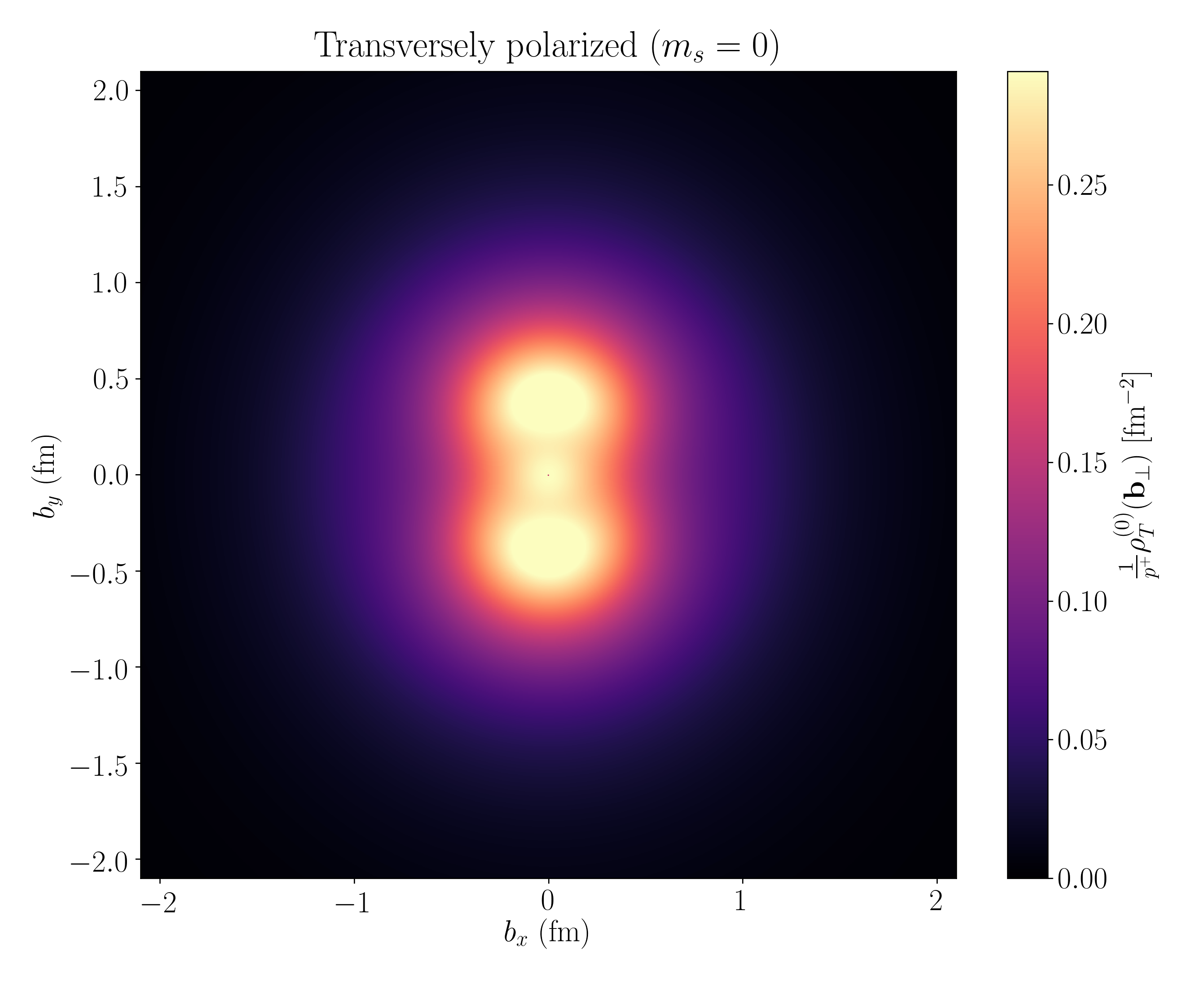}
  \includegraphics[width=0.49\textwidth]{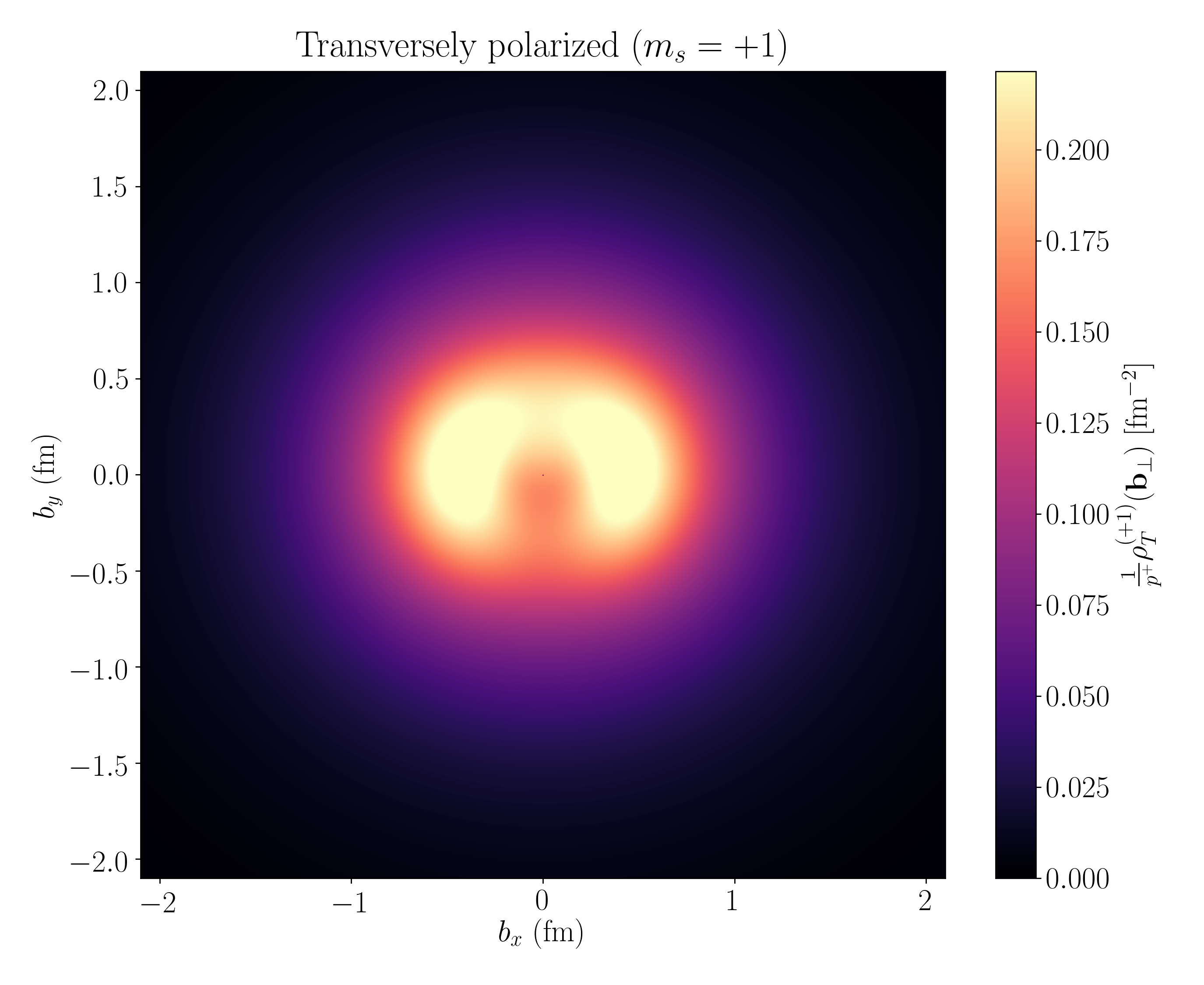}
  \caption{
    The $p^+$ density of the deuteron in various polarization states,
    with $p^+$ divided out to provide a boost-invariant density
    that is normalized to $1$.
    A left-handed coordinate system is used, so that the $z$ direction
    is into rather than out of the page.
    The panels are
    (top-left) helicity state with $\lambda=0$,
    (top-right) helicity state with $\lambda=+1$,
    (bottom-left) transverse polarization along $x$ axis with $m_s=0$,
    and
    (bottom-right) transverse polarization along $x$ axis with $m_s=+1$.
  }
  \label{fig:conv:p+}
\end{figure}

First, in Fig.~\ref{fig:conv:p+}, we present light front momentum
($p^+$) densities for both $\lambda=0$ and $\lambda=+1$ helicity states,
as well as for the $m_s=0$ and $m_s=+1$ transversely polarized states.
The $p^+$ densities obtained from this model are especially robust,
since they are obtained through the ``good'' component $T^{++}$ of the EMT.
They also provide the clearest, most transparent description of
the deuteron's structure.

A curious aspect of the $m_s=+1$ state is its deformation
towards the $+y$ direction.
This is a peculiarity of the use of light front coordinates,
and has been noticed for the deuteron's electric charge density previously
in Ref.~\cite{Carlson:2008zc,Lorce:2022jyi},
as well as in both the charge density~\cite{Burkardt:2002hr}
and $p^+$ density~\cite{Freese:2021qtb} of a transversely polarized proton.
In Ref.~\cite{Burkardt:2002hr},
this deformation was interpreted in terms of distortions created
by the point of view of an observer moving quickly towards the target.
However, no such reference frame has actually been chosen here.

The transverse deformations are likely due to a mix of different effects,
including the use of fixed $x^+$ rather than fixed $x^0$,
the fact that $x^-$ has been integrated out,
and that we are considering a density of $P^+$ rather than a density of $P^0$.
Note that there is more $P^3$ on one side of the axis of rotation than the other,
since the axis of rotation is along the $x$ axis,
and modulations in the $P^3$ density would be present in the $P^+$ density as well,
even in a three-dimensional instant form density in the rest frame.
\added{
There may also be modulation effects from Wigner-Melosh rotations connecting
states with light front spin and canonical spin,
as was observed in Refs.~\cite{Lorce:2020onh,Lorce:2022jyi,Chen:2022smg}
for spin-half systems.
}

Let us consider the static quantities associated with the momentum densities.
Starting with the radii, using Eq.~(\ref{eqn:p+:radius}),
we find the following radii for helicity states:
\begin{subequations}
  \begin{align}
    \langle b_\perp \rangle_{p^+}(\lambda=0)
    &=
    1.77~\mathrm{fm}
    \\
    \langle b_\perp \rangle_{p^+}(\lambda=0)
    &=
    1.69~\mathrm{fm}
    \\
    \overline{ \langle b_\perp \rangle_{p^+} }
    &=
    1.72~\mathrm{fm}
    \,.
  \end{align}
\end{subequations}
These results are roughly compatible with the known charge radius of the deuteron.
The Breit frame deuteron charge radius is $2.130$~fm~\cite{Sick:1998cvq},
which scaled down by $\sqrt{2/3}$ to give a rough estimate for a 2D charge radius,
gives $1.739$~fm.

For the transversely polarized states, we can calculate a quadrupole moment.
Using Eqs.~(\ref{eqn:quadrupole}) and (\ref{eqn:Amp}),
the light front quadrupole moment is found to be:
\begin{align}
  \mathcal{Q}_{\mathrm{LF}}
  =
  4
  \lim_{t\rightarrow0}
  \frac{\mathcal{A}_{-+}(t)}{t}
  =
  0.27~p^+\text{-}\mathrm{fm}
  \,,
\end{align}
which is surprisingly close to the empirical value of the electric quadrupole moment,
$0.2859$~$e$-fm~\cite{Code:1971zz,Bishop:1979zz,Ericson:1982ei}.

The pressure distributions are of special interest,
due to the amount of attention these have received in the
hadron physics community recently.
Unfortunately, the light cone convolution model is less trustworthy
for these quantities because they correspond to ``bad'' components
of the EMT, namely, $T^{ij}$.
An ideal situation would be to obtain $\mathcal{D}_{\lambda\lambda'}(t)$
from a manifestly covariant model.
Nonetheless, for illustration of the formalism,
we present the pressure distributions obtained from this model.

\begin{figure}
  \includegraphics[width=0.49\textwidth]{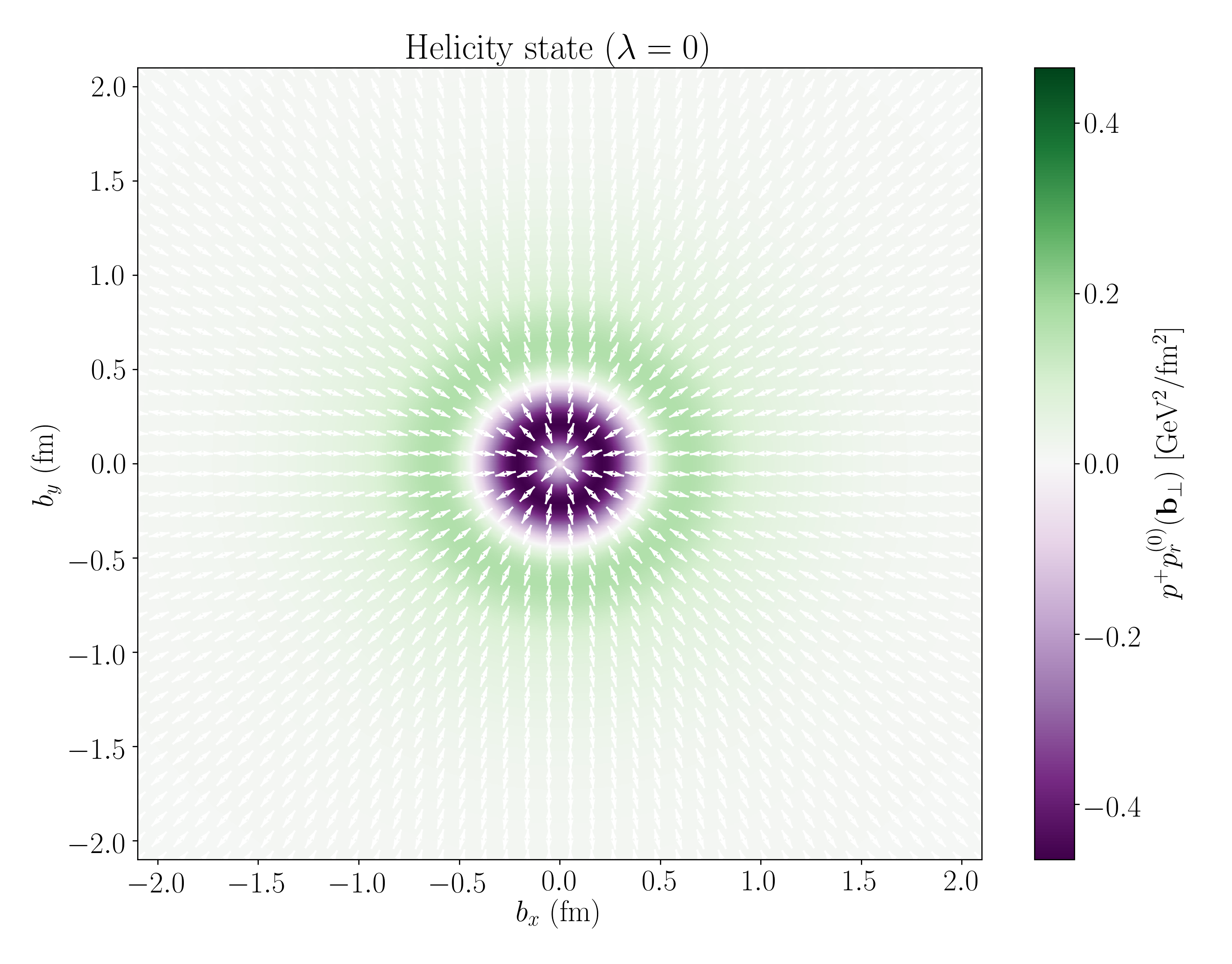}
  \includegraphics[width=0.49\textwidth]{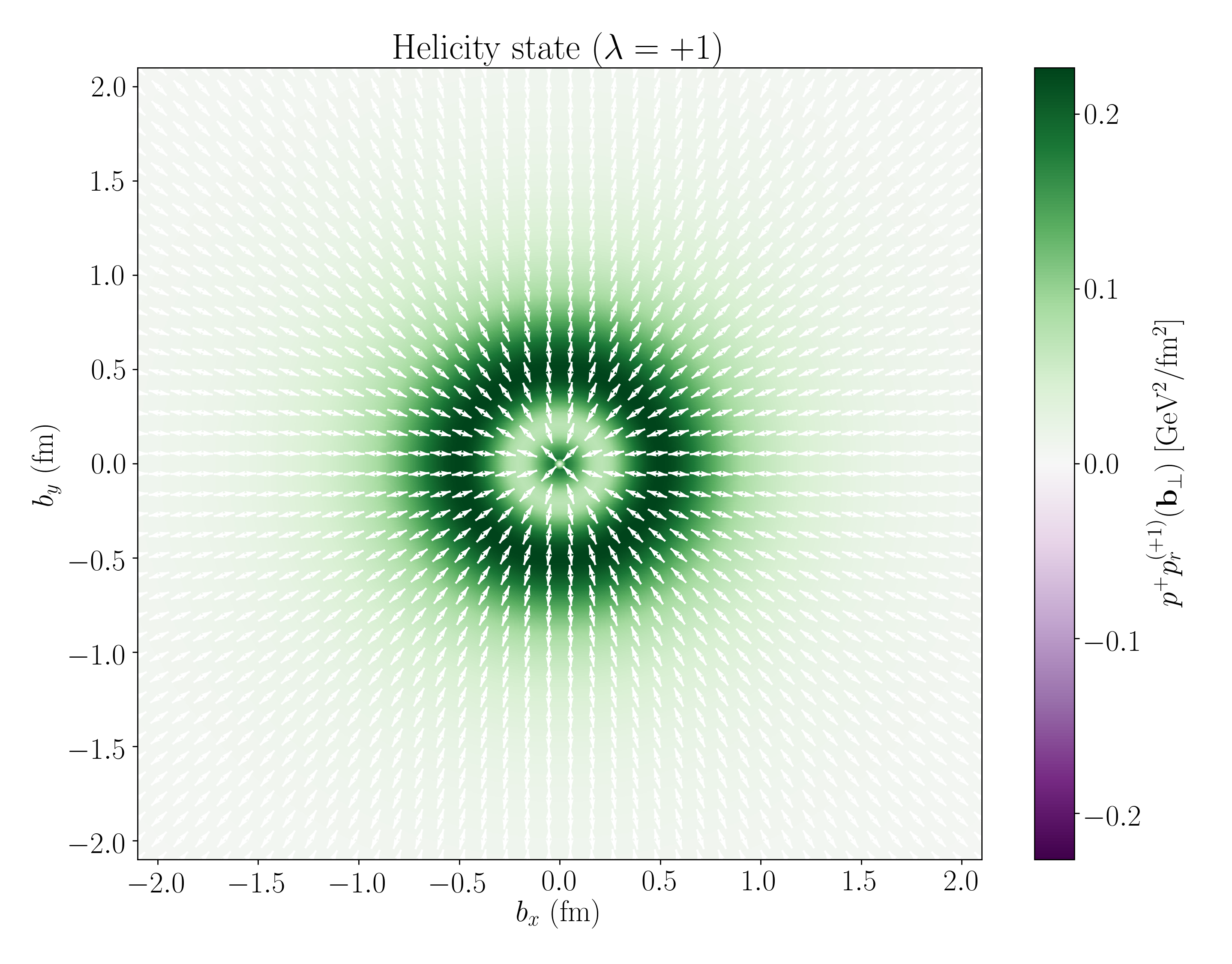}
  \includegraphics[width=0.49\textwidth]{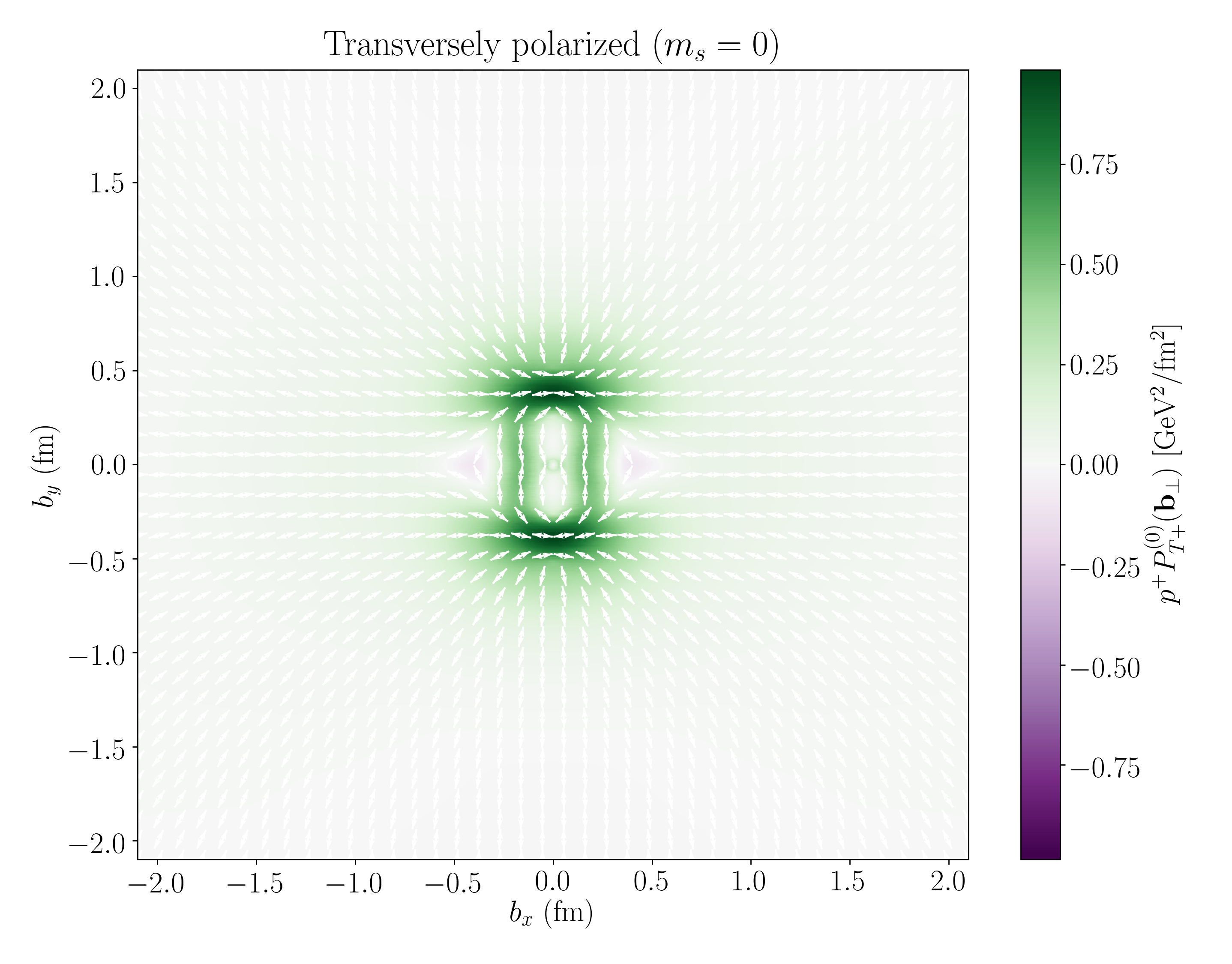}
  \includegraphics[width=0.49\textwidth]{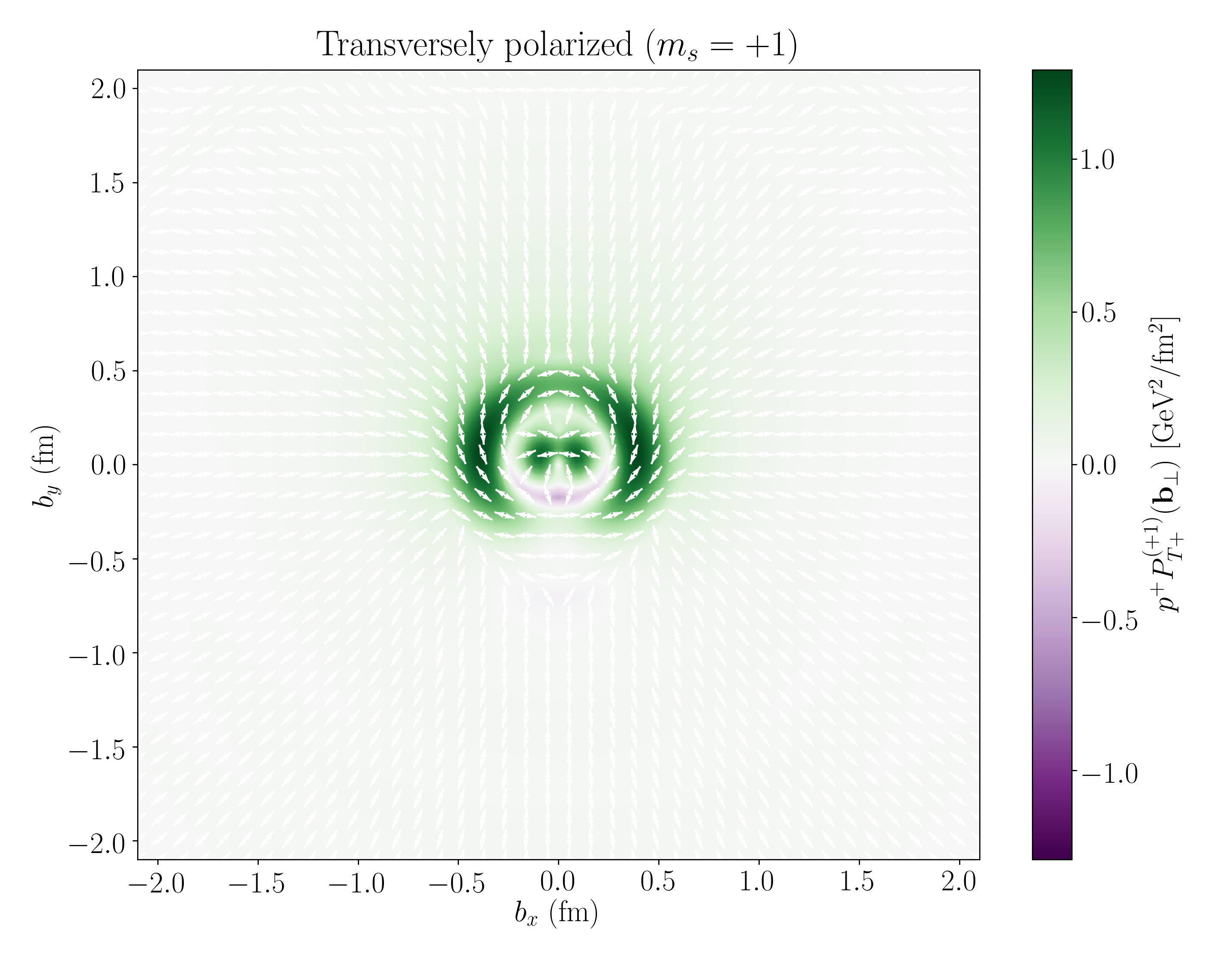}
  \caption{
    The radial (or $+$) eigenpressure of the deuteron in various polarization states,
    multiplied by $p^+$ to provide a boost-invariant density.
    See Eq.~(\ref{eqn:eigenpressure}) and the discussion around it
    for an explanation of the eigenpressures.
    The arrows indicate directions in which the pressure is acting,
    and are double-sided because pressures from both direction
    act with the same magnitude and result in a net zero force
    (see text for further elaboration).
    A left-handed coordinate system is used, so that the $z$ direction
    is into rather than out of the page.
    The panels are
    (top-left) helicity state with $\lambda=0$,
    (top-right) helicity state with $\lambda=+1$,
    (bottom-left) transverse polarization along $x$ axis with $m_s=0$,
    and
    (bottom-right) transverse polarization along $x$ axis with $m_s=+1$.
  }
  \label{fig:conv:stress}
\end{figure}

A selection of eigenpressures are presented in Fig.~\ref{fig:conv:stress},
with the selection limited to save space.
For helicity states, the radial pressure is selected,
and for transversely polarized states, the ``$+$'' eigenpressure
is selected according to Eq.~(\ref{eqn:eigenpressure}).
The color is selected to show magnitude and sign of the pressure,
and two-sided arrows to signify direction.

We feel it is important to reiterate the physical meaning of intrinsic pressure
and its sign in this context,
as was explained previously in Ref.~\cite{Freese:2021qtb}.
Since the deuteron is in equilibrium,
the expectation value of the force acting over any region of the transverse
plane is exactly zero.
By Gauss's theorem, this means that the integral of
$\mathbf{F}_\perp \cdot \hat{n}$ over the surface of any region must be zero.
The stresses encoded by the expectation value of $T^{ij}$ correspond to forces
acting on this region from all directions,
which sum to a net force of zero.
A positive pressure therefore does \emph{not} indicate a net repulsive
force from the center,
nor does a negative pressure signify a confining force towards the
center, as was claimed in Ref.~\cite{Burkert:2018bqq}.
A positive pressure means that particles in this region of space
are experiencing pushing forces \emph{from both directions},
and a negative pressure likewise means they are experiencing
pulling forces \emph{from both directions}.
For the radial eigenpressures (helicity states),
these directions are towards and away from the center of the deuteron,
while for transversely polarized states, the directions
are indicated by white arrows overlaid on the plot.

In fact, since the densities obtained in this formalism
correspond to stresses seen by transversely comoving observers,
the pressures are \emph{static} pressures or \emph{intrinsic} pressures,
and should be contrasted by \emph{dynamic} pressures
which include impulse imparted by flow or motion of the medium
(see for instance Chapter 4-3 of Ref.~\cite{Binder1943:fluid}).

It has been postulated throughout the literature~\cite{Polyakov:2018zvc,Lorce:2018egm,Freese:2021czn}
that the radial pressure should be positive as a stability condition.
Our result for the $\lambda=0$ radial pressure, in the top-left panel
of Fig.~\ref{fig:conv:stress}, violates this expectation.
Although the stability requirement is merely a conjecture lacking proof,
it is premature to declare our model result to be a counter-example,
owing to the possible shortcomings of a light cone convolution model.
For now, we consider the results here to be tentative and open to
replacement by results from a manifestly covariant calculation.

If we do however take the results in Fig.~\ref{fig:conv:stress} at face value,
they paint an interesting picture of the dynamics at play within the deuteron.
There appears to be a ring of roughly half a femtometer at which pressure
is more intense.
Within this ring, near the center, the pressure becomes negative
for the helicity zero state---specifically in the region where the
$p^+$ density is depleted (see Fig.~\ref{fig:conv:p+}).
The exact meaning of this negative pressure
(and its reality, given limitations of the model) is unclear.
One possibility is that the negative pressure corresponds to attractive
forces pulling particles inside the ring towards the ring,
and that the pressure remains negative because the pressure exerted by
other particles crowding the area is not present.

Let us consider static quantities associated with the comoving stress tensor.
First of all, the static D-terms for helicity states are:
\begin{subequations}
  \begin{align}
    \mathcal{D}_0(0)
    &=
    -24.33
    \,,
    \\
    \mathcal{D}_{\pm1}(0)
    &=
    -24.16
    \,.
  \end{align}
\end{subequations}
These values are large, negative, and nearly identical.
It is worth noting that negativity of $\mathcal{D}(0)$
has been frequently postulated~\cite{Perevalova:2016dln,Polyakov:2018zvc,Lorce:2018egm,Freese:2021czn}
as a \emph{looser} stability criterion than the radial pressure being positive,
and that our deuteron model at least satisfies this condition.
Next, we consider mechanical radii calculated according to
Eq.~(\ref{eqn:mechrad:D}):
\begin{subequations}
  \begin{align}
    \langle b_\perp \rangle_{\mathrm{mech}}(\lambda=0)
    &=
    2.39~\mathrm{fm}
    \\
    \langle b_\perp \rangle_{\mathrm{mech}}(\lambda=+1)
    &=
    1.06~\mathrm{fm}
    \\
    \overline{ \langle b_\perp \rangle_{\mathrm{mech}} }
    &=
    1.24~\mathrm{fm}
    \,.
  \end{align}
\end{subequations}
These results are surprising.
It is worth stressing, as discussed above,
that the ``average'' involves averaging the numerator and denominator
separately, rather than taking the mean of the three radii;
this is why the average mechanical radius is not close to the mean
of the three polarization states' radii.
In any case, the disparity between the radii is stark,
and can be understood clearly by looking at Fig.~\ref{fig:conv:stress}:
the negative presssure near the center of the helicity-zero state
greatly enhances its mechanical radius.


\section{Summary and outlook}
\label{sec:conclusion}

In this work, we obtained the two-dimensional light front densities
of momentum, angular momentum, and pressures within spin-one targets.
In contrast to the spin-half case, the densities have helicity dependence,
and the densities of transversely polarized spin-one hadrons can exhibit
quadrupole deformations that are related to the differences between the
helicity-one and helicity-zero densities.
All of these special properties of spin-one light front densities have
been illustrated with a light front convolution model of the deuteron.

Experimentally, the spatial densities for the deuteron could be extracted from data for coherent hard exclusive reactions on the deuteron.  These are challenging measurements, however, due to the steeper $t$-slopes of the coherent deuteron cross section compared to that of the nucleon. Current data is scarce: the HERMES collaboration has measured deeply virtual Compton scattering (DVCS) on the deuteron with both unpolarized~\cite{Airapetian:2009bm} and polarized targets~\cite{Airapetian:2010aa}, and Jefferson Lab (JLab) has more recent results for coherent $\pi^0$ electroproduction on the deuteron~\cite{Mazouz:2017skh}.  In the future, more data should be forthcoming from JLab~\cite{Armstrong:2017wfw,Munoz:PR12-22-006}, and especially the future electron-ion collider~\cite{Boer:2011fh,AbdulKhalek:2021gbh} with its dedicated far-forward detectors setup.  Accessing the gravitational form factors from these data is a non-trivial inverse problem, as they are related to Mellin moments of twist-2 vector generalized parton distributions (GPDs)~\cite{Cosyn:2019aio}, which are present in the amplitudes in the Compton form factors, being $x$-convolutions of the GPDs with a hard scattering coefficient.

In a following companion paper~\cite{Freese:2022ibw}, we apply the formalism developed here
to the photon as a special case.
A few minor modifications are made to accommodate the massless case,
but these result in simplifications of the formalism.
The photon is an especially pertinent target to consider,
since the employment of a light front formalism allows for its densities
to be calculated.

\begin{acknowledgments}
  The authors would like to thank Ian Clo\"et and Gerald Miller
  for enlightening discussions that helped contribute to this work.
  AF was supported by the U.S.\ Department of Energy
  Office of Science, Office of Nuclear Physics under Award Number
  DE-FG02-97ER-41014.
  WC was supported by the National Science Foundation under Award No.
2111442.
\end{acknowledgments}


\appendix

\section{Light front spin-one polarization vectors}
\label{app:polar}

This appendix uses the polarization vectors from Ref.~\cite{Berger:2001zb,Cano:2003ju,Cosyn:2018rdm},
but at $\xi=0$,
which is the case relevant to local operators such as the EMT.
Note that $\Delta^+ = -2\xi P^+$,
so having $\xi=0$ is equivalent to having $\Delta^+=0$,
and we take $\Delta^+ = (n\cdot\Delta) = 0$ throughout the paper
(including this appendix).

The polarization basis vectors are given explicitly by:
\begin{align}
  \varepsilon_0^\mu
  &=
  \frac{1}{M} \left(
  p^\mu - M^2\frac{n^\mu}{(P\cdot n)}
  \right)
  \\
  \varepsilon'^\mu_0
  &=
  \frac{1}{M} \left(
  p'^\mu - M^2\frac{n^\mu}{(P\cdot n)}
  \right)
  \\
  \varepsilon^\mu_1
  &=
  \frac{1}{\sqrt{-t}} \left(
  \Delta^\mu + t\frac{n^\mu}{2(P\cdot n)}
  \right)
  \\
  \varepsilon'^\mu_1
  &=
  \frac{1}{\sqrt{-t}} \left(
  \Delta^\mu - t\frac{n^\mu}{2(P\cdot n)}
  \right)
  \\
  \varepsilon^\mu_2
  &=
  \varepsilon'^\mu_2
  =
  -
  \frac{1}{\sqrt{-t}}
  \frac{
    \epsilon^{\mu}_{\phantom{\mu}\nu\alpha\beta} P^\nu \Delta^\alpha n^\beta
  }{(P\cdot n)}
  \,,
\end{align}
where the Levi-Civita symbol is normalized to satisfy
$\varepsilon^{0123} = +1$.
These polarization basis vectors satisfy the following orthogonality
and normalization relations:
\begin{align}
  \varepsilon_i\cdot\varepsilon_j
  &=
  \varepsilon'_i\cdot\varepsilon'_j
  =
  -\delta_{ij}
  \\
  \varepsilon_i\cdot p
  &=
  \varepsilon'_i\cdot p'
  =
  0
  \,.
\end{align}
The positive and negative helicity vectors are defined via~\cite{Cosyn:2018rdm}\footnote{
  Note that Refs.~\cite{Berger:2001zb,Cano:2003ju} take $\phi_\Delta=\pi$ and define $\epsilon_1^\mu, \epsilon_2^\mu$ (and primed equivalent vectors) with an opposite sign.
}:
\begin{align}
  \label{eqn:helicity:pm}
  \varepsilon_{\pm}
  &=
  \mp e^{\pm i\phi_\Delta}
  \frac{ \varepsilon_1 \pm i\varepsilon_2 }{\sqrt{2}}
  \,,
\end{align}
and equivalently for the primed four-vectors,
where $\phi_\Delta$ is the azimuthal angle of
the momentum transfer $\boldsymbol{\Delta}_\perp$
with respect to a fixed $\hat{x}$ axis.
The positive and negative helicity vectors satisfy:
\begin{align}
  \varepsilon_+^*\cdot\varepsilon_+
  &=
  \varepsilon'^*_-\cdot\varepsilon'_-
  =
  -1
  \\
  \varepsilon_+^*\cdot\varepsilon_-
  &=
  \varepsilon'^*_-\cdot\varepsilon'_+
  =
  0
  \\
  \varepsilon_0\cdot p
  &=
  \varepsilon_+\cdot p
  =
  \varepsilon_-\cdot p
  =
  0
  \\
  \varepsilon'_0\cdot p'
  &=
  \varepsilon'_+\cdot p'
  =
  \varepsilon'_-\cdot p'
  =
  0
  \,.
\end{align}
For $\varepsilon_{\pm}$ (and their primed counterparts)
specifically:
\begin{align}
  \varepsilon_+\cdot n
  &=
  \varepsilon'_+\cdot n
  =
  \varepsilon_-\cdot n
  =
  \varepsilon'_-\cdot n
  =
  0
  \,.
\end{align}
For massless spin-one particles such as the photon,
using $\varepsilon_\pm$ as the polarization vectors
thus amounts to using light cone gauge.

Several helpful explicit four-products include:
\begin{align}
  (\varepsilon_0\cdot\Delta)
  &=
  \frac{-t}{2M}
  \\
  (\varepsilon'_0\cdot\Delta)
  &=
  \frac{+t}{2M}
  \\
  (\varepsilon_1\cdot\Delta)
  &=
  (\varepsilon'_1\cdot\Delta)
  =
  -\sqrt{-t}
  \\
  (\varepsilon_2\cdot\Delta)
  &=
  (\varepsilon'_2\cdot\Delta)
  =
  0
  \,,
\end{align}
and several explicit outer products include:
\begin{align}
  \frac{1}{2}
  \varepsilon_0^{\{\mu} \varepsilon_0^{\prime\nu\}}
  &=
  \frac{1}{M^2}
  \left( P^\mu P^\nu - \frac{1}{4} \Delta^\mu \Delta^\nu \right)
  -
  \frac{ n^{\{\mu} P^{\nu\}} }{(P\cdot n)}
  +
  \frac{M^2 n^\mu n^\nu}{(P\cdot n)^2}
  \\
  \frac{1}{2}
  \varepsilon_1^{\{\mu} \varepsilon_1^{\prime\nu\}}
  &=
  -
  \frac{\Delta^\mu \Delta^\nu}{t}
  +
  \frac{t}{4}
  \frac{n^\mu n^\nu}{(P\cdot n)^2}
  \\
  \frac{1}{2}
  \varepsilon_2^{\{\mu} \varepsilon_2^{\prime\nu\}}
  &=
  -
  g^{\mu\nu}
  -
  \left(1 - \frac{t}{4M^2}\right)
  \frac{M^2 n^\mu n^\nu}{(P\cdot n)^2}
  +
  \frac{ n^{\{\mu} P^{\nu\}} }{(P\cdot n)}
  +
  \frac{\Delta^\mu \Delta^\nu}{t}
\end{align}


\section{Explicit EMT matrix elements}
\label{app:elements}

In this Appendix, we give explicit evaluations of all the
helicity amplitudes
in
Eq.~(\ref{eqn:emt:spin1}) in terms of the $\mathcal{G}_{1-6}(t)$ form factors
in Eq.~(\ref{eqn:emt:massive}),
using all combinations of the polarization vectors in Appendix~\ref{app:polar}.

Firstly, for the $\mathcal{A}$ results:
\begin{align}
  \mathcal{A}_{00}
  &=
  \left( 1 + \frac{t}{2M^2} \right)
  \mathcal{G}_1(t)
  - \frac{t}{4M^2} \Big(
  2 \mathcal{G}_5(t) + \mathcal{G}_6(t)
  \Big)
  - \frac{t^2}{8M^4} \mathcal{G}_2(t)
  \\
  \mathcal{A}_{++}
  &=
  \mathcal{A}_{--}
  =
  \mathcal{G}_1(t) - \frac{t}{4M^2} \mathcal{G}_2(t)
  \\
  \mathcal{A}_{0+}
  &=
  \mathcal{A}_{-0}
  =
  - \mathcal{A}^*_{+0}
  =
  - \mathcal{A}^*_{0-}
  =
  -
  \frac{ \sqrt{-t} }{ \sqrt{2} M }
  \left\{
    \mathcal{G}_1(t) - \frac{1}{2} \mathcal{G}_5(t)
    - \frac{t}{4M^2} \mathcal{G}_2(t)
    \right\}
  e^{i\phi_\Delta}
  \\
  \label{eqn:Amp}
  \mathcal{A}_{-+}
  &=
  \mathcal{A}^*_{+-}
  =
  \frac{t}{4M^2} \mathcal{G}_2(t)
  e^{2i\phi_\Delta}
  \,.
\end{align}
Next, for $\mathcal{J}(t)$:
\begin{align}
  \mathcal{J}_{00}
  &=
  0
  \\
  \mathcal{J}_{++}
  &=
  -
  \mathcal{J}_{--}
  =
  \frac{1}{2} \mathcal{G}_5(t)
  \\
  \mathcal{J}_{0+}
  &=
  - \mathcal{J}_{-0}
  =
  -\mathcal{J}^*_{+0}
  =
  \mathcal{J}^*_{0-}
  =
  -
  \frac{\sqrt{-t}}{4\sqrt{2}M}
  \Big\{
    \mathcal{G}_5(t) + \mathcal{G}_6(t)
    \Big\}
  e^{i\phi_\Delta}
  \\
  \mathcal{J}_{-+}
  &=
  \mathcal{J}^*_{+-}
  =
  0
  \,.
\end{align}
Next, for $\mathcal{D}(t)$:
\begin{align}
  \mathcal{D}_{00}
  &=
  \left( 1 + \frac{t}{2M^2} \right)
  \mathcal{G}_3(t)
  - \frac{t}{4M^2} \mathcal{G}_6(t)
  - \frac{t^2}{8M^2} \mathcal{G}_4(t)
  \\
  \mathcal{D}_{++}
  &=
  \mathcal{D}_{--}
  =
  \mathcal{G}_3(t) - \mathcal{G}_6(t) - \frac{t}{4M^2} \mathcal{G}_4(t)
  \\
  \mathcal{D}_{0+}
  &=
  \mathcal{D}_{-0}
  =
  - \mathcal{D}^*_{+0}
  =
  - \mathcal{D}^*_{0-}
  =
  -
  \frac{ \sqrt{-t} }{ \sqrt{2} M }
  \left\{
    \mathcal{G}_3(t) - \frac{1}{2} \mathcal{G}_6(t)
    - \frac{t}{4M^2} \mathcal{G}_4(t)
    \right\}
  e^{i\phi_\Delta}
  \\
  \mathcal{D}_{-+}
  &=
  \mathcal{D}^*_{+-}
  =
  \frac{t}{4M^2} \mathcal{G}_4(t)
  e^{2i\phi_\Delta}
  \,.
\end{align}
Next, for $\mathcal{E}(t)$:
\begin{align}
  \mathcal{E}_{00}
  &=
  \frac{t}{2} \Big( \mathcal{G}_5(t) + \mathcal{G}_6(t) \Big)
  \\
  \mathcal{E}_{++}
  &=
  \mathcal{E}_{--}
  =
  \frac{t}{4} \Big( \mathcal{G}_5(t) - \mathcal{G}_6(t) \Big)
  \\
  \mathcal{E}_{0+}
  &=
  \mathcal{E}_{-0}
  =
  - \mathcal{E}^*_{+0}
  =
  - \mathcal{E}^*_{0-}
  =
  -
  \frac{\sqrt{-t} M}{2\sqrt{2}}
  \left\{
    \left(1 + \frac{t}{4M^2}\right) \mathcal{G}_5(t)
    +
    \frac{t}{4M^2} \mathcal{G}_6(t)
    \right\}
  e^{i\phi_\Delta}
  \\
  \mathcal{E}_{-+}
  &=
  \mathcal{E}^*_{+-}
  =
  -
  \frac{t}{4} \Big( \mathcal{G}_5(t) + \mathcal{G}_6(t) \Big)
  e^{2i\phi_\Delta}
  \,.
\end{align}
Next, for $\mathcal{H}(t)$:
\begin{align}
  \mathcal{H}_{00}
  &=
  -\frac{tM^2}{2} \mathcal{G}_6(t)
  \\
  \mathcal{H}_{++}
  &=
  \mathcal{H}_{--}
  =
  \frac{tM^2}{4}
  \left( 1 - \frac{t}{2M^2} \right) \mathcal{G}_6(t)
  \\
  \mathcal{H}_{0+}
  &=
  \mathcal{H}_{-0}
  =
  - \mathcal{H}^*_{+0}
  =
  - \mathcal{H}^*_{0-}
  =
  \frac{t\sqrt{-t}M}{4\sqrt{2}} \mathcal{G}_6(t)
  e^{i\phi_\Delta}
  \\
  \mathcal{H}_{-+}
  &=
  \mathcal{H}^*_{+-}
  =
  \frac{tM^2}{4} \mathcal{G}_6(t)
  e^{2i\phi_\Delta}
  \,.
\end{align}
Lastly, for $\mathcal{K}(t)$:
\begin{align}
  \mathcal{K}_{00}
  &=
  0
  \\
  \mathcal{K}_{++}
  &=
  -
  \mathcal{K}_{--}
  =
  \frac{t}{8} \mathcal{G}_6(t)
  \\
  \mathcal{K}_{0+}
  &=
  \mathcal{K}_{-0}
  =
  - \mathcal{K}^*_{+0}
  =
  - \mathcal{K}^*_{0-}
  =
  -
  \frac{\sqrt{-t}M}{4\sqrt{2}}
  \mathcal{G}_6(t)
  e^{i\phi_\Delta}
  \\
  \mathcal{K}_{-+}
  &=
  \mathcal{K}^*_{+-}
  =
  0
  \,.
\end{align}

We also state the expressions for the non-Galilean effective form factors for general polarization
\begin{align}
    \sum_{\lambda,\lambda'} \rho(\lambda,\lambda')\, \mathcal{E}_{\lambda' \lambda} =&
    \left(\frac{2}{3}+T_{LL}\right) \frac{t}{4}\left(\mathcal{G}_5(t)-\mathcal{G}_6(t) \right) + \left(\frac{1}{3}-T_{LL}\right) \frac{t}{2}\left(\mathcal{G}_5(t)+\mathcal{G}_6(t) \right)\nonumber\\
    & \qquad +i S_T\sin(\phi_S-\phi_t)\, \frac{M\sqrt{-t}}{2} \left[\left(1+\frac{t}{4M^2}\right)\mathcal{G}_5(t) + \frac{t}{4M^2}\mathcal{G}_6(t) \right]
    \nonumber\\
    & \qquad -T_{TT}\cos(2\phi_{T_T}-2\phi_t)\, \frac{t}{4}(\mathcal{G}_5(t)+\mathcal{G}_6(t))\\[1em]
    \sum_{\lambda,\lambda'} \rho(\lambda,\lambda')\, \mathcal{H}_{\lambda' \lambda} =&
    \left[ \left(\frac{2}{3}+T_{LL}\right)\frac{M^2t}{4}\left(1-\frac{t}{2M^2} \right) -\left(\frac{1}{3}-T_{LL}\right)\frac{M^2t}{2}\right.\\
    &\qquad\left.-iS_T\sin(\phi_S-\phi_t)\,\frac{Mt\sqrt{-t}}{4}
    +T_{TT}\cos(2\phi_{T_T}-2\phi_t)\,\frac{M^2t}{4}
    \right] \mathcal{G}_6(t)
    \\[1em]
    \sum_{\lambda,\lambda'} \rho(\lambda,\lambda')\, \mathcal{K}_{\lambda' \lambda} =& \left( S_L \frac{t}{8} + i T_{LT}\sin(\phi_{T_L}-\phi_t) \,\frac{M\sqrt{-t}}{2}\right) \mathcal{G}_6(t)
\end{align}

\section{Spin-1 density matrix}
\label{app:rho}

The density matrix $\rho(\lambda,\lambda')$ of a spin-1 system is a 3$\times$3 Hermitian matrix 
with unit trace, $\sum_{\lambda,\lambda'}\rho(\lambda,\lambda')=1$. In the rest frame (RF) of the spin-1 system it can be
specified in a basis of single-particle states $|\bm{p} = \bm0; \lambda\rangle$,
where the momentum is zero and the spin is quantized along the $z$-axis, with 
spin projection $\lambda = (-1, 0, 1)$. The density matrix can be parametrized in the form
\cite{Leader:2001gr}
\begin{equation} \label{eq:dens_nonrel}
\rho \equiv \rho(\lambda,\lambda') = 
\frac{1}{3}+\frac{1}{2}S_i\mathscr{S}_i + 
T_{ij}\mathscr{T}_{ij}\,.
\end{equation}
Here, $\mathscr{S}_{i}$ are the $3\times 3$ matrices describing the spin 
operators
in the spin--1 representation for $\bm e_\pm = \mp \frac{1}{\sqrt{2}}(\bm e_x \pm i \bm e_y); \bm e_0 = \bm e_z$ ,
\begin{equation}
\mathscr{S}_x 
= \frac{1}{\sqrt{2}} \left( \begin{array}{rrr} 0 & \phantom{-}1 & \phantom{-}0 
\\ 1 & 0 & 1 \\ 0 & 1 & 0 \end{array} \right),
\hspace{2em}
\mathscr{S}_y = 
\frac{i}{\sqrt{2}} \left( \begin{array}{rrr} 0 & -1 & \phantom{-}0 
\\ 1 & 0 & -1 \\ 0 & 1 & 0 \end{array} \right),
\hspace{2em}
\mathscr{S}_z = 
\left( \begin{array}{rrr} 1 & \phantom{-}0 & \phantom{-}0 
\\ 0 & 0 & 0 \\ 0 & 0 & -1 \end{array} \right) ,
\end{equation}
and their symmetric traceless rank-2 tensors
\begin{equation}
\mathscr{T}_{ij} = \frac{1}{2}(\mathscr{S}_i\mathscr{S}_j+\mathscr{S}_j\mathscr{S}_i)-\frac{2}{3}\delta_{ij},
\end{equation}

and $i,j = (x, y, z)$ denote the Cartesian components. The parameters in 
Eq.~(\ref{eq:dens_nonrel})
are a 3-dimensional vector $S_i$ and a traceless symmetric tensor $T_{ij}$. They 
coincide, respectively,
with the expectation value of the spin operators and their traceless tensor 
products
\begin{subequations}
\begin{align}
&S_i =\textrm{Tr} [\rho \hat{\mathscr{S}}_i] = 
\langle\hat{\mathscr{S}}_i\rangle,
\label{eq:nonrel_spin_vector}
\\[1ex]
&T_{ij} = 
\textrm{Tr} \left[ \rho 
\hat{\mathscr{T}}_{ij} \right]
 =
\langle \hat{\mathscr{T}}_{ij} \rangle.
\label{eq:nonrel_spin_tensor}
\end{align}
\end{subequations}

In the rest frame of a particle, the covariant spin-1 density matrix $\rho^{\beta \alpha}[RF]$ can be introduced as

\begin{equation} \label{eq:rho_cov_RF}
    \rho^{\beta \alpha}[RF] = \sum_{\lambda,\lambda'} \rho(\lambda,\lambda') \epsilon^\beta(k,\lambda) \epsilon^{*\alpha} (k,\lambda') = \frac{1}{3}\left(-g^{\beta \alpha} + \frac{k^\beta k^\alpha}{M^2}\right) \; -\; \frac{i}{2M}\epsilon^{\beta \alpha s[\text{RF}] k} \; - \; t^{\beta\alpha}[\text{RF}]\,,
\end{equation}
where $k^\mu = (M,0,0,0)$. In the rest frame, the spin vector $s^\mu[\text{RF}]$ and tensor $t^{\beta\alpha}[\text{RF}]$ only have spatial components, which are identical to the spin parameters appearing in Eq.~(\ref{eq:dens_nonrel}). The spin tensor is traceless.  In formulas:
\begin{subequations}
\label{eq:rf_spin_parameters}
\begin{align}
    &s^i[\text{RF}] \equiv S_i, 
    &s^0[\text{RF}] = 0, \\
    &t^{ij}[\text{RF}] \equiv T_{ij},
    &t^{0\alpha}[\text{RF}] = t^{\beta 0}[\text{RF}] =0, \qquad \tensor{t}{^{\mu}_\mu}[\text{RF}]=0.
\end{align}
\end{subequations}
It is advantageous to consider the following (2+1)D (transverse, longitudinal) decomposition of the  rest frame spin vector and tensor: 
\begin{subequations}
\label{eq:transverse_spin}
\begin{align}
    & S = (S_x,S_y,S_x) \equiv S_L\,(0,0,1) + S_T\,(\cos \phi_S,\sin \phi_S,0)\\[2em]
    &T = \left( \begin{array}{cc|c} \frac{T_{xx}-T_{yy}}{2} - \frac{T_{zz}}{2} & T_{xy} & T_{xz}\\ T_{xy} & -\frac{T_{xx}-T_{yy}}{2} - \frac{T_{zz}}{2} & T_{yz} \\[0.3em] \hline  T_{xz} & T_{yz} & T_{zz}\end{array} \right)  \equiv \left( \begin{array}{c|c} \bm T_{TT} - \frac{T_{LL}}{2}1_{2\times 2} & \bm T_{LT} \\[0.3em]\hline \bm T_{LT} & T_{LL} \end{array} \right),\nonumber\\[1em] 
    &\bm T_{LT} = (T_{xz},T_{yz})\equiv  T_{LT} \,(\cos \phi_{T_L},\sin \phi_{T_L}),\\[1em]
    &\bm T_{TT} =\left( \begin{array}{cc} \frac{T_{xx}-T_{yy}}{2} & T_{xy}\\ T_{xy} & -\frac{T_{xx}-T_{yy}}{2}\end{array} \right) 
    \equiv 
    \frac{T_{TT}}{2}\left( \begin{array}{cc} \cos 2\phi_{T_T} & \sin 2\phi_{T_T}\\ \sin 2\phi_{T_T} & -\cos 2\phi_{T_T} \end{array} \right)\,,
\end{align}
\end{subequations}
where $\bm T_{TT}$ is symmetric and traceless in transverse coordinates.

In cases where one considers an outer product of two polarization vectors with specific helicity values $\lambda,\lambda'$, one can obtain the value of $\epsilon^\beta(k,\lambda) \epsilon^{*\alpha} (k,\lambda')$ by making the following substitutions in Eq.~(\ref{eq:rho_cov_RF}) for the unpolarized, vector and tensor polarized parts, see Eqs.~(\ref{eq:nonrel_spin_vector} and (\ref{eq:nonrel_spin_tensor}):
\begin{subequations}
\begin{align}
& 1 \rightarrow \delta_{\lambda,\lambda'},\\
& S_i \rightarrow  \langle \lambda' | \mathscr{S}_i | \lambda \rangle,\\
& T_{ij} \rightarrow \langle \lambda' | \mathscr{T}_{ij} | \lambda \rangle.
\end{align}
\end{subequations}
Similar statements apply for the polarization parameters introduced in Eq.~(\ref{eq:transverse_spin}).

For the case of particles with non-zero three-momentum we can introduce the density matrix by applying Lorentz boosts to the polarization four vectors in Eq.~(\ref{eq:rho_cov_RF})~\cite{Leader:2001gr,Cosyn:2019aio,Cosyn:2020kwu}.  Different choices of standard boosts transforming the rest frame particle to the moving one result in different expressions for the polarization vectors (connected by the so-called Melosh rotations)~\cite{Melosh:1974cu,Polyzou:2012ut}.  As we consider the EMT on the light front, we only consider light front boosts $\Lambda_\text{LF}(p)$ here:  
\begin{multline} \label{eq:rho_forward}
        \rho^{\beta \alpha}(p) = \sum_{\lambda,\lambda'} \rho(\lambda,\lambda') \tensor{\Lambda_\text{LF}(p)}{^\beta_\mu} \epsilon^\mu(k,\lambda) \tensor{\Lambda_\text{LF}(p)}{^\alpha_\nu} \epsilon^{*\nu} (k,\lambda')  \\ = 
        \sum_{\lambda,\lambda'} \rho(\lambda,\lambda') \epsilon^\beta(p,\lambda)\epsilon^{*\alpha} (p,\lambda') = \frac{1}{3}\left(-g^{\beta \alpha} + \frac{p^\beta p^\alpha}{M^2}\right) \; -\; \frac{i}{2M}\epsilon^{\beta \alpha s p} \; - \; t^{\beta\alpha}
\end{multline}
where the covariant spin vector and tensor are introduced as
\begin{subequations}
\begin{align}
    &p^\mu = \tensor{\Lambda_\text{LF}(p)}{^\mu_\nu} k^\mu
    &s^\mu = \tensor{\Lambda_\text{LF}(p)}{^\mu_\nu}\, s^\nu[\text{RF}], \quad (sp) = 0,\\
    &t^{\mu\nu} = \tensor{\Lambda_\text{LF}(p)}{^\mu_\rho}\,\tensor{\Lambda_\text{LF}(p)}{^\nu_\sigma}\, t^{\rho\sigma}[\text{RF}], 
    &t^{\beta \alpha} p_\alpha = p_\beta t^{\beta \alpha}=0, \qquad \tensor{t}{^\mu_\mu}=0.
\end{align}
\end{subequations}

To compute the spatial densities of the EMT, we need off-diagonal bilinears of spin-1 polarization four vectors.  We therefore need the expression for the off-diagonal covariant density matrix.  We write this expression using the averaged spin vector $\bar{s}$ and tensor $\bar{t}$~\cite{Lorce:2017isp}, which are obtained by boosting the rest frame ones of Eq.~(\ref{eq:rf_spin_parameters}) with the average momentum $P=\frac{p+p'}{2}$:
\begin{subequations}
\begin{align}
    &\bar{s}^\mu = \tensor{\Lambda_\text{LF}(P)}{^\mu_\nu} s^\nu[\text{RF}],\\
    &\bar{t}^{\mu\nu} = \tensor{\Lambda_\text{LF}(P)}{^\mu_\rho}\tensor{\Lambda_\text{LF}(P)}{^\nu_\sigma} t^{\rho\sigma}[\text{RF}].
\end{align}
\end{subequations}

The off-diagonal density matrix then becomes
\begin{align}\label{eqn:rho_off}
    \rho^{\beta\alpha}_\text{LF}(p,p') = \sum_{\lambda,\lambda'}\, &\rho(\lambda,\lambda')  \, \tensor{\Lambda_\text{LF}(p)}{^\beta_\mu}\epsilon^\mu(k,\lambda)  \, \tensor{\Lambda_\text{LF}(p')}{^\alpha_\nu}\epsilon^{*\nu} (k,\lambda')= \nonumber\\
    \frac{1}{3}&\left[-g^{\beta\alpha} 
    +\frac{P^\beta P^\alpha}{M^2}
    -\frac{\Delta^\beta \Delta^\alpha}{4M^2}
    +\frac{P^{[\beta}\Delta^{\alpha]}}{2M^2}
    +\frac{ \Delta^{[\beta}n^{\alpha]}+\xi\,\Delta^{\{\beta}n^{\alpha\}}}{\left(1-\xi^2\right)(P\cdot n)}
    +\frac{\Delta^2}{2(1-\xi^2)} \frac{n^\beta n^\alpha }{(P\cdot n)^2}\right]\nonumber\\
    +\frac{i}{2MD}&\Bigg\{
    \epsilon^{\beta\alpha\bar{s}P}
    - \left(D-1\right)(\bar{s} \cdot n)\frac{M^2}{(P\cdot n)^2} \epsilon^{\beta\alpha+P}
    -(D-1)\frac{P^{[\beta}\epsilon^{\alpha]n\bar{s}P}}{(P\cdot n)}
    +\frac{D}{2}\frac{\Delta^{\{\beta}\epsilon^{\alpha\} n\bar{s}P}}{(P\cdot n)}\nonumber\\
    &\quad -D(D-1)\frac{M^2}{\left(1-\xi^2\right)(P\cdot n)^2}\left(n^{[\beta}\epsilon^{\alpha]n\bar{s}P} - \xi \,
    n^{\{\beta}\epsilon^{\alpha\}n\bar{s}P} \right)
   \nonumber\\[1em]
    &\quad+\frac{\epsilon^{n\Delta\bar{s}P}}{2\left(1-\xi^2\right)(P\cdot n)^2}
    \left[-2D(D-1)\frac{M^2}{(P\cdot n)} n^\beta n^\alpha 
    \right. \nonumber\\[1em]
    &\qquad \left.
    - (D-1) \left(P^{\{\beta}n^{\alpha\}}+\xi\,P^{[\beta}n^{\alpha]}\right)
    +\frac{D}{2}\left(\Delta^{[\beta}n^{\alpha]}+\xi\,\Delta^{\{\beta}n^{\alpha\}} \right)
    \right]\nonumber\\[1em] 
    &
    \quad-\frac{n^{\{\beta}\epsilon^{\alpha\}\Delta\bar{s}P} - \xi \,
    n^{[\beta}\epsilon^{\alpha]\Delta\bar{s}P}}{2\left(1-\xi^2\right)(P\cdot n)}
    +(D-1)\frac{(\bar{s}\cdot n)}{(P\cdot n)}M^2\frac{n^{\{\beta}\epsilon^{\alpha\}\Delta n P} - \xi \,
    n^{[\beta}\epsilon^{\alpha]\Delta n P}}{2\left(1-\xi^2\right)(P\cdot n)^2}
    \Bigg\}\nonumber\\[1em]
    &\quad -\bar{t}^{\beta\alpha}
    +\frac{\Delta^{[\beta}\bar{t}^{\alpha] n} }{2(P\cdot n)}
    -\frac{n^{[\beta}\bar{t}^{\alpha]\Delta}-\xi\,n^{\{\beta}\bar{t}^{\alpha\}\Delta} }{2(1-\xi^2)(P\cdot n)}
    +\bar{t}^{nn}\frac{\Delta^\beta\Delta^\alpha}{4(P\cdot n)^2}\nonumber\\
    &\quad
    -\bar{t}^{n\Delta}\frac{n^{\{\beta}\Delta^{\alpha\}}-\xi n^{[\beta}\Delta^{\alpha]}}{4(1-\xi^2)(P\cdot n)^2}
    +\bar{t}^{\Delta\Delta} \frac{n^\beta n^\alpha}{4\left(1-\xi^2\right)(P\cdot n)^2},
\end{align}

where
\begin{subequations}
\begin{align}
    &D= \sqrt{1-\frac{t}{4M^2}},\\
    &a^{\{\mu}b^{\nu\}}=a^\mu b^\nu + a^\nu b^\mu,
    &a^{[\mu}b^{\nu]}=a^\mu b^\nu - a^\nu b^\mu.
\end{align}
\end{subequations}


\bibliography{references.bib}

\end{document}